\documentclass[aps,pra, twocolumn]{revtex4-1}
\usepackage{amsthm}

\usepackage[utf8]{inputenc}
\usepackage[sc,osf]{mathpazo}

\usepackage{graphicx,subfigure}
\usepackage{bm}
\usepackage{makeidx} 
\usepackage{amsmath}
\usepackage{amssymb}
\usepackage{color}
\usepackage{amsthm}
\usepackage{hyperref}
\usepackage{cleveref}
\usepackage{diagbox}
\usepackage{physics}
\hypersetup{colorlinks, citecolor=magenta, filecolor=blue, linkcolor=blue, urlcolor=green}
\usepackage[normalem]{ulem}
\usepackage{float}
\usepackage{multirow}

\newcolumntype{P}[1]{>{\centering\arraybackslash}p{#1}}
\begin{document}

\title{ Measurement-based Multipartite Entanglement Inflation}

\author{Pritam Halder$^{1}$, Shiladitya Mal$^{1, 2}$, Aditi Sen(De)$^{1}$}
	
	\affiliation{$^1$ Harish-Chandra Research Institute and HBNI, Chhatnag Road, Jhunsi, Allahabad - 211019, India\\
	$^2$ Department of Physics and Center for Quantum Frontiers of Research and Technology (QFort),
National Cheng Kung University, Tainan 701, Taiwan
	}

\begin{abstract}
Generating entanglement between more parties is one of the central tasks and challenges in the backdrop of building quantum technologies. Here we propose a measurement-based protocol for producing multipartite entangled states which can be later fed into some network for realizing  suitable quantum protocols. We consider weak entangling measurement on two parties as  the basic unit of operation to create entanglement between more parties starting from an entangled state with a lesser number of parties and auxiliary systems in the form of a single-qubit or entangled state itself. We call the introduced expansion procedure, ``multipartite entanglement inflation''. In the context of inflating bipartite entanglement to more number of parties, surprisingly, maximally entangled states as inputs turn out to be worse than that of the non-maximally entangled states, Haar uniformly generated pure states having a moderate amount of entanglement and the Werner state with a certain threshold noise. We also report that the average multipartite entanglement created from the initial Greenberger Horne Zeilinger- and the W- class states are almost  same. Interestingly, we also observe that for  Haar uniformly generated pure states,   unentangled auxiliary systems are sometimes more  advantageous  than the protocol with multiple copies of the initial entangled states.

\end{abstract}

\maketitle

\section{Introduction}
\label{sec_intro}

In the second revolution of quantum technologies, multipartite entangled states are shown to play a crucial role in designing computational tasks  which include  measurement-based quantum computers \cite{hans'01, qcomp1, qcomp2, qcomp3}, distributed quantum computation \cite{beals'13}, and communication protocols like teleportation \cite{tele, teleexp, tele1, tele2, tele3}, dense coding networks \cite{dc, dcexp, dc1, dc2, dc2, dc3, dc4, dc5, dc6}, quantum secret sharing   \cite{ekert'91, key1, key2, key3, key4, key5, hillery'06}, conference key agreement \cite{xu'14}. 
Moreover, the patterns of multisite entanglement turn out to be important in addressing  fundamental questions like a quantum phase transition, dynamical phase transition in many-body physics \cite{qmanybody, qmanybody1, qmanybody2, dqpt, qmanybody3}. 

The entire development in the technological front demands a systematic creation and detection of multipartite entangled states which is one of the current challenges in quantum information science. To accomplish it, several procedures were proposed --   the desired  state is created  in a single location and is then distributed to others by means of teleportation \cite{linden'99}; initial entanglement is generated which is then manipulated to obtain  the desired state
\cite{pirker'18}; two-qubit quantum gates are employed between atoms to finally create a multiatomic entangled state in optical lattices \cite{jaksch'99}.
Another prominent method in setting up quantum networks through noisy channels is the discovery of a quantum repeater  \cite{hans'98} based on entanglement swapping \cite{zukowski'93, swapping2, swapping3, swapping4} and distillation protocol \cite{bennett'96}. It was later shown that by removing the step of distillation,  multipartite projective measurements on one part  of the multiple copies of noisy entangled states, referred to as a star network, can also lead to a multipartite entangled state \cite{swapping5, network2, swapping6, epping'16}. 
 From the perspective of circuit models, generating any multisite entangled state can be described by sequential adaptive applications of an universal set 
 of single- and two-qubit  quantum gates \cite{fusion}. This technique led to  protocols like quantum state expansion, where cat-states are  generated by employing a chain of two-party controlled-NOT (CNOT) gates on a product state of single qubits \cite{lee'05}, fusion mechanism, in which Fredkin gate or CNOT  along with Toffoli gates are used to create multipartite entangled  states \cite{fusion1, fusion2}. 

In the present work, we propose a measurement-based protocol for generating multipartite entangled  states where  weak entangling measurement on two parties serves as the basic unit of operation \cite{unsharp1, unsharp1a}.  
In the literature, the potential of this kind of joint measurement has been less studied compared to the exploration of various features of multipartite entangled states. However, it was realized that  the positive-operator valued measurements can give improvements in different quantum information tasks like  state-tomography, detection of entanglement, violation of Bell inequalities, discrimination of states, randomness generation \cite{adv1, adv2, adv3, adv4, adv5, adv6}. 
Moreover, it is interesting to note that although projective joint measurements  can not be taken as the basic operation for the purpose of creating entanglement between more number of parties starting from a  lesser number of parties, weak entangling measurement has the potential to fulfill this task. We call the proposed expansion procedure,
 the ``multipartite entanglement inflation'' process. 

As a building block, weak (or unsharp) Bell state measurement \cite{unsharp4, unsharp5}  is applied on part of a bipartite state and an auxiliary qubit for the purpose of inflating entanglement to a higher number of parties (see Fig. \ref{Fig:schematic}). 
The performance of the proposed protocol is measured by  the maximal possible  genuine multipartite entanglement   generated in the resulting states. In particular, multipartite entanglement  is quantified via  generalized geometric measure (GGM) \cite{tele3, biswas'14} and negativity monogamy score\cite{dhar'17}  while the maximization is performed over the  parameters involved in  the auxiliary systems and in the weak measurement.  We report that non-maximally entangled two-qubit states can create higher genuine multipartite entangled (GME) states  than that of the maximally entangled ones. 
More precisely, we find that for a given sharpness parameter, there always exists a unique two-qubit entangled state which can yield the maximum genuine multipartite entanglement in the output state,  obtained recursively for arbitrary rounds. Such an observation is also confirmed by considering Haar uniformly simulated two-qubit pure states  \cite{bengtsson'06} and the Werner state \cite{werner'89} as inputs. For the latter case, we demonstrate that for a fixed value of weakness parameter in the measurement, there is a threshold noise value where  the negativity monogamy score is maximized.  
Starting from the tripartite entangled state, we find that for a given genuine multipartite entanglement, generalized Greenberger  Horne Zeilinger (GHZ) state \cite{GHZ} can produce the maximal GGM in the output state compared to any three-qubit pure states belonging to both the GHZ- and the W-class \cite{dur'00} after the first round of measurement. 

We then extend the inflation protocol by considering many  copies of the initial  bipartite states as auxiliary systems. Interestingly, we observe that although in this scenario, the initial state  contains a higher amount of entanglement on average compared to the former process with qubit-auxiliary states, it is not always successful to create high GME states. 
 

The paper is organized as follows. In Sec. \ref{sec_network}, we introduce the multipartite entanglement inflation procedure with product auxiliary states. Taking bipartite states as initials, the recursion relation of resulting state after the arbitrary number of rounds of entangling measurement is derived and the patterns of genuine multipartite entanglement are analyzed in Sec. \ref{sec_genproduct} while the similar method is extended for the initial tripartite state in Sec. \ref{sec:multitomulti}. Instead of product single-qubit auxiliary states, when we use the entangled resource state as the auxiliary state itself, the method and the multipartite entanglement content of the output state are discussed in Sec. \ref{sec_pic2}. The concluding remarks are discussed in Sec. \ref{sec:conclu}. 

\section{Multipartite entanglement inflation Process with qubit Auxiliary}
\label{sec_network}

\begin{figure}[ht]
	\includegraphics[width=0.9\linewidth]{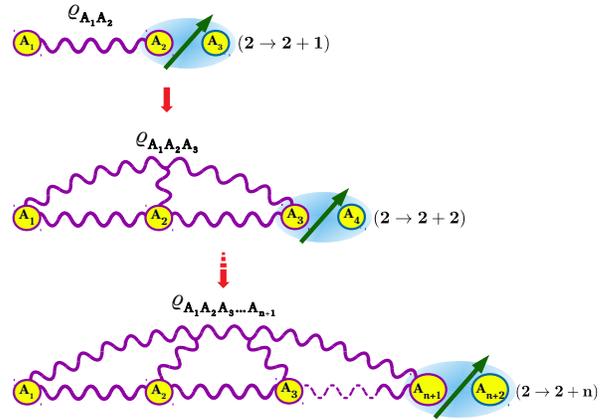}
	\caption{ Schematic diagram of multipartite inflation. In the first round, denoted by \(2 \rightarrow 2 +1\), starting from a two-party state, \(\varrho_{A_1A_2}\) and a single qubit auxiliary state, \(\varrho_{A_3}\), a weak entangling measurement (blue patch) is performed in one part of the two-party state, \(A_2\) and on \(A_3\). Similar method continues up to round \(n\), when \((2+n)\)-party state is created via \(2 \rightarrow 2 +n\) inflation process. In each round, we maximize GME by optimizing state  and sharpness parameters.  }
	\label{Fig:schematic}
\end{figure}

Let us describe the protocol to obtain genuine multpartite entangled states, which we refer as entanglement inflation process. The resources required in this scenario are -- (1) an entangled state of \(m\) parties; (2)  auxiliary qubits and  the operations performed is the set of weak joint measurements, \(\{\sqrt{M_k (\lambda)}\}\) (with  $\sum_{k}M_k=\mathbb{I}$,) having control parameter \(\lambda\). The weak measurement performed on one of the parties of an entangled states and the auxiliary qubit plays a crucial role to extend multipartite entangled state and the  parameters in the measurement as well as auxiliary systems can control the content of multipartite entanglement of the output state. 

Let us illustrate the scenario for the initial shared bipartite state,
$\varrho_{A_1 A_2}$ and an auxiliary system, $\varrho_{A_3} (\theta, \phi)$, with \(\theta\), and \(\phi\) being the state parameters, resulting to an initial state,  $\rho^{1} =\varrho_{A_1 A_2}\otimes\varrho_{A_3}(\theta, \phi)$. After the weak  measurement $\sqrt{M_{k}}$ performed jointly on  parties $A_2$ and $A_3$, the resulting tripartite state  becomes 
\begin{eqnarray}
\rho^{1}_{k} \equiv \varrho_{A_1 A_2 A_3}= \frac{\sqrt{M_{k} (\lambda)} \rho^{1} \sqrt{M_{k} (\lambda)}^{\dagger}}{\mbox{Tr}\left(\sqrt{M_{k} (\lambda)} \rho^{1} \sqrt{M_{k} (\lambda)}^{\dagger}\right)},
\end{eqnarray}
 where the subscript $k$ corresponds to the outcome $k$ of the measurement while the superscript $1$ denotes the first round of the measurement (see Fig. \ref{Fig:schematic}). \(\lambda =1\) represents the projective joint measurement for which the output state becomes the biseparable state,  \(\varrho_{A_1} \otimes \varrho_{A_2 A_3}\) while for other values of \(\lambda\), there is a possibility to create a tripartite entangled state. Our aim is to maximize the amount of multipartite entanglement of the above state, quantified by a suitable measure, \(\mathcal{E}\),  i.e., 
 \begin{eqnarray}
 \mathcal{E}^{2 \rightarrow 2 +1}_c= \max_{\lambda, \theta, \phi}  \mathcal{E}(\rho^{1}_{k}),
 \label{eq:PBarrow}
 \end{eqnarray}
 where maximization is performed over the sharpness parameters of the weak measurement as well as the  state parameters of the auxiliary system. In this paper, generalized geometric measure (GGM) \cite{tele3, biswas'14}, \(\mathcal{G}\),   and negativity monogamy score \cite{coffman'00, dhar'17}, \(\delta_{\mathcal{N}}\), are used to measure multipartite entanglement content of the resulting pure and mixed states respectively. 
This is the first step of the protocol towards inflating GME state, and is denoted by \(2 \rightarrow 2 +1\). 
 
 After $n$ rounds of measurements, denoted by \(2 \rightarrow 2+n\) where \(2\) denotes the number of parties in the initial resource state,  a genuine ($2+n$)-party entangled state, $\rho^{n}_{k}=\varrho_{A_1 A_2 A_3 A_4 ... A_{2+n}}= \frac{\sqrt{M_{k}} \rho^{n} \sqrt{M_{k}}^{\dagger}}{\mbox{Tr}\left(\sqrt{M_{k}} \rho^{n} \sqrt{M_{k}}^{\dagger}\right)}$ is created. 
Therefore, starting from a two qubit entangled state, and optimizing multipartite entanglement,  \(\mathcal{E}^{2 \rightarrow 2+n}_c = \max_{\lambda, \theta_1, \phi_1, \ldots}  \mathcal{E}(\rho^{n}_{k})\) over parameters of auxiliary states, \(\theta_1, \phi_1, \ldots\) and measurement, \(\lambda\),  we  can, in principle,  expand GME state via weak measurement as shown in Fig. \ref{Fig:schematic}. Notice that the state prepared in this process is for the specific outcome, thereby showing its probabilistic nature. The design of the protocol has to be made in such a way that either all the outcomes  leads to the  output state having almost equal genuine multipartite entanglement or one has to optimize over the probability of the measurement to obtain  high GME states.  In this paper, we will show that our choice of measurement reflects the former situation.  

\begin{figure}[ht]
	\includegraphics[width=0.9\linewidth]{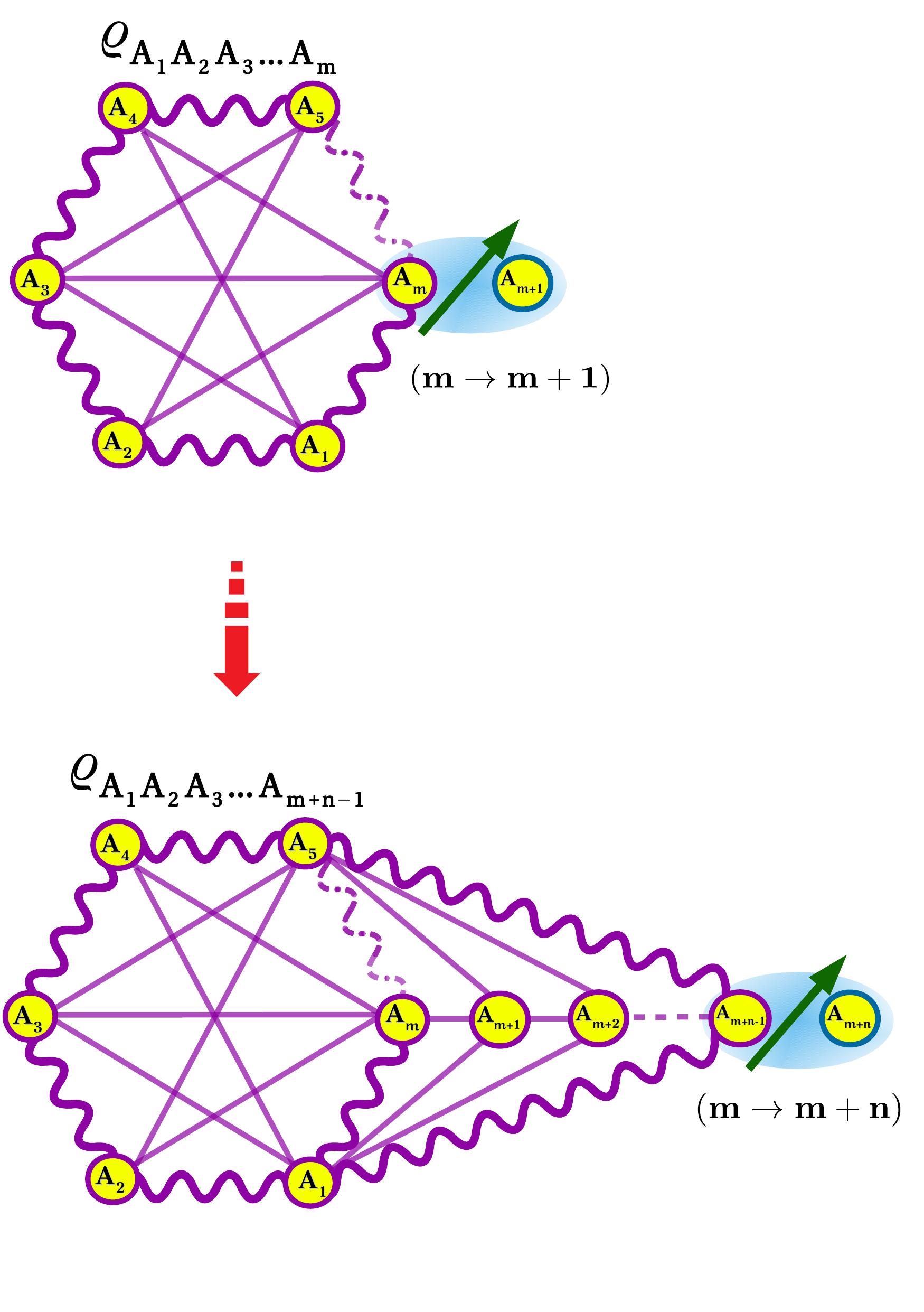}
	\caption{Schematic diagram to create multipartite entangled states. It is similar to Fig. \ref{Fig:schematic} except the  \(m\)-party resource state, \(\varrho_{A_1 \ldots A_m}\). In this picture,   after \(n\) rounds, \((n+m)\)-party state is produced and hence   the process can be denoted as \(m \rightarrow m+n\). }
	\label{Fig:mschematic}
\end{figure}
Instead of bipartite entangled state, if we start with \(m\)-party entangled state and maximize entanglement with respect to the set of measurements and auxiliary state parameters as before, we may result a \(m+n\)-party GME state (see Fig. \ref{Fig:mschematic}). In this work, we consider two- and three-party initial states and show expansion of GME states having \(2(3)+n\) parties. Moreover,  in this picture, the auxiliary systems  are considered as a product form, and so we call this process as product-based (PB) inflation. We will compare this scenario with entangled auxiliary systems  which we refer as entanglement-based (EB) inflation in the succeeding section.
In both the PB and EB methods, it is interesting to find out whether any  class of multipartite states, important for different quantum information processing tasks,  can be generated  by suitably choosing an initial multiparty entangled state with less number of parties, an auxiliary state and a weak measurement.  Such studies may reveal  an interesting connection between the multipartite initial state, the weak measurement and the resulting state.


%


\section{Genuine Multisite Entanglement Production with Two-party entangled resource (\(2\rightarrow 2+n\))}
\label{sec_genproduct}

Let us exhibit whether the procedure described in the previous section can indeed generate  genuine multipartite entangled  states  or not. To answer it, we choose  maximally, non-maximally, Haar uniformly generated entangled  pure states and  the Werner state as the initial states. To show the usefulness of the method, it is necessary to depict it for  a set  of weak measurement, which in our case, reads as
\begin{eqnarray}
M_k(\lambda)&=&f^{1}_{k}(\lambda)\ket{\psi^+}\bra{\psi^+}+f^{2}_{k}(\lambda)\ket{\psi^-}\bra{\psi^-}\\ \nonumber &+&f^{3}_{k}(\lambda)\ket{\phi^+}\bra{\phi^+}+f^{4}_{k}(\lambda)\ket{\phi^-}\bra{\phi^-}, 
\end{eqnarray}
$\big\{\ket{\psi^+}, \ket{\psi^-}, \ket{\phi^+}, \ket{\phi^-}\big\}$ is the Bell basis. 
Let us choose these $f^{i}_{k}$'s in such a way that these operators become
\begin{eqnarray}
  \nonumber 	&&M_1=\lambda\ket{\psi^+}\bra{\psi^+}+(1-\lambda)\frac{\mathbb{I}_2}{4}\\&& \nonumber
	M_2=\lambda\ket{\psi^-}\bra{\psi^-}+(1-\lambda)\frac{\mathbb{I}_2}{4}\\&& \nonumber M_3=\lambda\ket{\phi^+}\bra{\phi^+}+(1-\lambda)\frac{\mathbb{I}_2}{4}\\&& M_4=\lambda\ket{\phi^-}\bra{\phi^-}+(1-\lambda)\frac{\mathbb{I}_2}{4},
	\label{eq:weakM}
\end{eqnarray}
with \(\lambda\) being the tuning parameter, and \(f^{1}_{k}(\lambda) = \sqrt{\frac{1+3\lambda}{4}}\) and \(f^2_k (\lambda) = f^3_k (\lambda) = f^4_k (\lambda)= \sqrt{\frac{1-\lambda}{4}}\).  As mentioned before, we will now maximize GGM and negativity monogamy score with respect to \(\lambda\) and auxiliary state parameters. 

\subsection{Maximally entangled state as initial state} 
\label{sec:pb}

Let us first describe in details the process of inflation when the initial shared state is the maximally entangled state, $\ket{\phi^+}$. The produced multipartite state after \(n\) rounds obtained via recursion method can then be extended to any other shared resource state. 
At the first round, the initial state  reads as $\ket{\phi^+}_{A_1 A_2}\otimes{\ket{\chi_1^{+}}}_{A_3}$ where \(\ket{\chi_i^{+}}=\alpha_i\ket{0}+\beta_i\ket{1}\)  and \(\alpha_i=\cos \frac{\theta_i}{2} \;;\;\beta_i=e^{i\phi_i}\sin \frac{\theta_i}{2}\),  with the subscripts, \(i=1, 2, \ldots, n\), represent the number of rounds. 
After performing the  first round of weak measurements in Eq. (\ref{eq:weakM}), for each outcome, \(\sqrt{M_k}\) (\(k=1,2,3,4\)), the corresponding four output tripartite state takes the form as
\begin{eqnarray*}
\ket{\Psi_{1}^{1}}&=&\frac{1}{2}\{\sqrt{1+3\lambda}\ket{\xi_1^{+}}\ket{\psi^+} + \sqrt{1-\lambda}(\ket{\xi_1^{-}}\ket{\psi^-}\\&+& \ket{\chi_1^{+}}\ket{\phi^+} + \ket{\chi_1^{-}}\ket{\phi^-})\}\\\\\ket{\Psi_{2}^{1}}&=&\frac{1}{2}\{\sqrt{1+3\lambda}\ket{\xi_1^{-}}\ket{\psi^-} + \sqrt{1-\lambda}(\ket{\xi_1^{+}}\ket{\psi^+}\\&+& \ket{\chi_1^{+}}\ket{\phi^+} + \ket{\chi_1^{-}}\ket{\phi^-})\}\\\\\ket{\Psi_{3}^{1}}&=&\frac{1}{2}\{\sqrt{1+3\lambda}\ket{\chi_1^{+}}\ket{\phi^+} + \sqrt{1-\lambda}(\ket{\xi_1^{+}}\ket{\psi^+}\\&+& \ket{\xi_1^{-}}\ket{\psi^-}+\ket{\chi_1^{-}}\ket{\phi^-})\}\\\\\ket{\Psi_{4}^{1}}&=&\frac{1}{2}\{\sqrt{1+3\lambda}\ket{\chi_1^{-}}\ket{\phi^-} + \sqrt{1-\lambda}(\ket{\xi_1^{+}}\ket{\psi^+}\\&+& \ket{\xi_1^{-}}\ket{\psi^-}+\ket{\chi_1^{+}}\ket{\phi^+})\},  
\end{eqnarray*} 
where \(\ket{\chi_n^{-}}=\alpha_n\ket{0}-\beta_n\ket{1}, \, \,  \ket{\xi_n^{\pm}}=\beta_n\ket{0}\pm\alpha_n\ket{1}\). 
On the other hand, if $\sqrt{M_1}$ clicks in the second round, the resulting four-party state becomes
\begin{eqnarray*}
\ket{\Psi_{1}^{2}}&=&\frac{1}{16 \sqrt{p_1^2}}\bigg[\bigg\{\sqrt{(1-\lambda)(1+3\lambda)}(\ket{\chi_1^{+}}\ket{\xi_2^{+}}\\&+&\ket{\chi_1^{-}}\ket{\xi_2^{-}}+\ket{\xi_1^{-}}\ket{\chi_2^{-}})\\&+&(1+3\lambda)\ket{\xi_1^{+}}\ket{\chi_2^{+}}\bigg\}\ket{\psi^+}\\&+&\bigg\{(1-\lambda)(\ket{\chi_1^{+}}\ket{\xi_2^{-}}+\ket{\chi_1^{-}}\ket{\xi_2^{+}}-\ket{\xi_1^{-}}\ket{\chi_2^{+}})\\&-&\sqrt{(1-\lambda)(1+3\lambda)}\ket{\xi_1^{+}}\ket{\chi_2^{-}}\bigg\}\ket{\psi^-}\\&+&\bigg\{(1-\lambda)(\ket{\chi_1^{+}}\ket{\chi_2^{+}}+\ket{\chi_1^{-}}\ket{\chi_2^{-}}+\ket{\xi_1^{-}}\ket{\xi_2^{-}})\\&+&\sqrt{(1-\lambda)(1+3\lambda)}\ket{\xi_1^{+}}\ket{\xi_2^{+}}\bigg\}\ket{\phi^+}\\&+&\bigg\{(1-\lambda)(\ket{\chi_1^{+}}\ket{\chi_2^{-}}+\ket{\chi_1^{-}}\ket{\chi_2^{+}}-\ket{\xi_1^{-}}\ket{\xi_2^{+}})\\&-&\sqrt{(1-\lambda)(1+3\lambda)}\ket{\xi_1^{+}}\ket{\xi_2^{-}}\bigg\}\ket{\phi^-}\bigg].
\end{eqnarray*}
For the outcome \(\sqrt{M_1}\),  the structure of the generated  three- and four-party state can help us to write the \((2+n)\)-party output state created after the round \(n\) with \(k\) being the outcome of the measurement, as
\begin{eqnarray}
\label{nth_state}
\ket{\Psi_{k}^{n}}&=&\frac{1}{4^{n} \sqrt{p_k^n}}\ket{R^n}=\frac{1}{\sqrt{p_k^n}}\ket{Z_k^n},
\end{eqnarray}
where
\begin{equation}
    \ket{R^n} = \bigg[ \ket{a^n}\ket{\psi^+}+\ket{b^n}\ket{\psi^-}+\ket{c^n}\ket{\phi^+}+\ket{d^n}\ket{\phi^-}\bigg].
    \label{recursion}
\end{equation}
and $p_k^n$ is the probability of $k$th outcome in the $n$th round. Here, for $n\geq2$,
\begin{eqnarray}
\nonumber \ket{a^n}&=& m_1^n \big[\ket{a^{n-1}}\ket{\chi_n^{+}}+\ket{b^{n-1}}\ket{\chi_n^{-}}\nonumber\\
&+&\ket{c^{n-1}}\ket{\xi_n^{+}}+\ket{d^{n-1}}\ket{\xi_n^{-}}\big], \nonumber\\ 
 \ket{b^n}&=& m_2^n \big[-\ket{a^{n-1}}\ket{\chi_n^{-}}-\ket{b^{n-1}}\ket{\chi_n^{+}} \nonumber \\
&+&\ket{c^{n-1}}\ket{\xi_n^{-}}+\ket{d^{n-1}}\ket{\xi_n^{+}}\big],\nonumber \\ 
\ket{c^n}&=& m_3^n \big[\ket{a^{n-1}}\ket{\xi_n^{+}}+\ket{b^{n-1}}\ket{\xi_n^{-}} \nonumber\\
&+&\ket{c^{n-1}}\ket{\chi_n^{+}}+\ket{d^{n-1}}\ket{\chi_n^{-}}\big], \nonumber\\ 
 \ket{d^n}&=& m_4^n \big[-\ket{a^{n-1}}\ket{\xi_n^{-}}-\ket{b^{n-1}}\ket{\xi_n^{+}} \nonumber\\
 &+&\ket{c^{n-1}}\ket{\chi_n^{-}}+\ket{d^{n-1}}\ket{\chi_n^{+}}\big],
 \label{eq:abcd}
\end{eqnarray}
with \(\sqrt{M_1}\) being the outcome,  
\begin{eqnarray*}
\ket{a^1}&=&\sqrt{1+3\lambda}\ket{\xi_1^+}\ ;\ \ket{b^1}=\sqrt{1-\lambda}\ket{\xi_1^-}\\\ket{c^1}&=&\sqrt{1-\lambda}\ket{\chi_1^+}\ ;\ \ket{d^1}=\sqrt{1-\lambda}\ket{\chi_1^-}.
\end{eqnarray*}
Here \(m_k^n\)s depend on the outcome of the measurement.  For example, if \(\sqrt{M_k}\) clicks , \(m_k^n = \sqrt{1 + 3 \lambda}\) while \(m_l^n = \sqrt{1 - \lambda}\) with \(l \neq k\). 
Due to the symmetry in the measurement and state, we find that in this situation,  multiparty entanglement  does not depend on the outcome of the measurement and so it is enough to maximize the GGM of the above states with respect to \(\lambda\) and \(\theta_n\)s, \(\phi_n\)s. 

Although the probabilities of clicking four different measurements are, in general, not equal, the GGM values remain same irrespective of the outcome of the measurement, thereby making the calculation of the probability unimportant (for relevant  discussion on probabilities, see Appendix). 

\begin{figure}[ht]
	\includegraphics[width=0.9\linewidth]{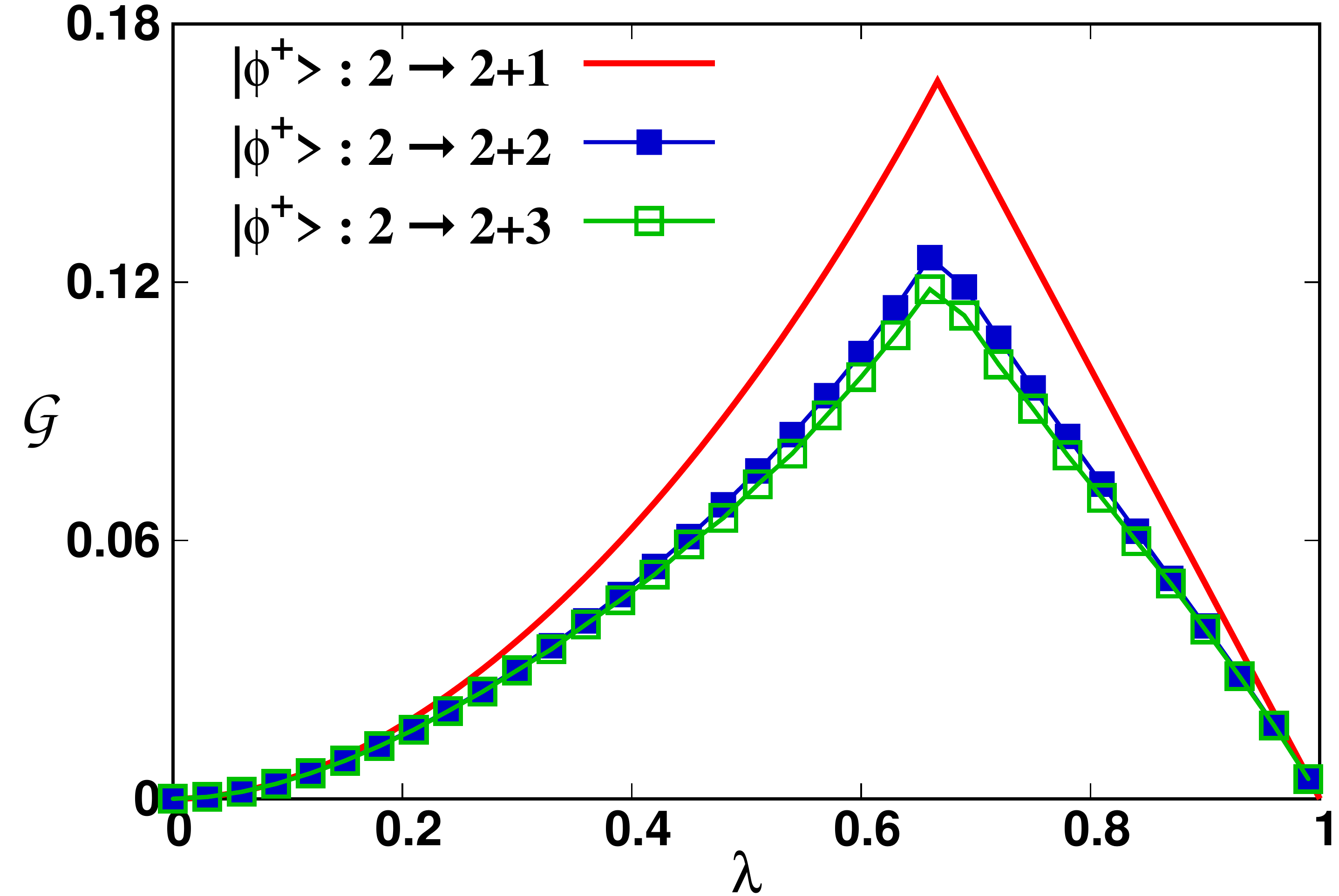}
	\caption{\emph{Maximally entangled state as resource.} GGM, \(\mathcal{G}\), (ordinate) vs. controlled parameters, \(\lambda\)  (abscissa) in  the weak measurement  in Eq. (\ref{eq:weakM}).  The maximally entangled state is initially shared and  arbitrary qubits are taken as auxiliary systems. After  first (solid line), second (filled squares) and third (hollow squares)  rounds,\(\mathcal{G}\) is obtained after maximizing over \(\theta_i, \phi_i\)s of the auxiliary states. In the entire paper, \(\mathcal{G}\) represents GGM after maximizing over auxiliary state parameters and \(\mathcal{G}_c\) denotes GGM after optimizing both state parameters and \(\lambda\). 
	In all the rounds, $\mathcal{G}$ is maximum at  
	$\lambda_c=\frac{2}{3}$, thereby giving \(\mathcal{G}_c\) although it is decreasing with the increase of the number of rounds. Both the axes are dimensionless. }
	\label{Fig:3}
\end{figure}

%
Let us now examine the behavior of GGM for the resulting state with the tuning parameter \(\lambda\) of the measurement after optimizing over input auxiliary systems in each round as depicted in  Fig.  \ref{Fig:3}. From Eq. (\ref{eq:weakM}), it is clear that \(\mathcal{G}\) vanishes both at \(\lambda =0\) and \(\lambda=1\). We find that GGM increases with \(\lambda\) till   $\lambda_c=\frac{2}{3}$ while it starts decreasing after that. 
%
In this respect, we also notice that after the first round, when \(\lambda < \lambda_c\), the maximum eigenvalue required to evaluate \(\mathcal{G}\) comes from the third party while \(\lambda >\lambda_c\),  the maximum eigenvalue of the reduced first-party state contributes, i.e., 
\begin{eqnarray}
&\mathcal{G}& = 1 - e_{A_3} = \frac{1}{2} \nonumber \\
&   - &\frac{\sqrt{2}}{4}\sqrt{(1-\lambda)\left(1+\lambda+\sqrt{(1-\lambda)(1+3\lambda)}\right)}, \, \, \lambda < \lambda_c, \ \nonumber \\
    &=& 1 - e_{A_1} = 1- \frac{1}{2}(1+\lambda), \, \, \lambda > \lambda_c.
\end{eqnarray}
Moreover, in this tripartite scenario,  GGM is independent of $\theta_1$, $\phi_1$ and  critical value of GGM, $\mathcal{G}_c=\frac{1}{6}$, corresponding to $\lambda_c=\frac{2}{3}$ where \(e_{A_3}\) and \(e_{A_1}\) coincide. For notational simplicity, we remove the superscript from the notation of  \(\mathcal{G}_c\) as seen in Eq. (\ref{eq:PBarrow}). 
In fact, all the four resulting states corresponding to other measurement outcomes lead to the same value of GGM due to symmetry.
From the second round onwards, we perform optimization over all the $\{\theta_i,  \phi_i\}$ to obtain the maximum values of GGM at each values of $\lambda$, since eigenvalues which contribute to GGM depends on $\theta_i$, $\phi_i$s.
Interestingly, we observe that for a fixed value of \(\lambda\),  although \(\mathcal{G}\) decreases in each round, \(\lambda_c\) remains same in all the rounds, and \(\mathcal{G}_c >0\),  thereby confirming genuine multipartite entanglement in all the output states via this process. We will later show the behavior of GGM by varying the number of rounds. 

Before moving to the other resource state, let us compute the tangle \cite{dur'00, coffman'00},
  \(  \delta_{\mathcal{C}^2} = \mathcal{C}^2 (\rho_{A_1 : A_2 A_3}) - \sum_{i=2}^3 \mathcal{C}^2  (\rho_{A_1 A_i})\)
of the tripartite state, where \(\mathcal{C}\) is the concurrence \cite{hill'97, wootters'98}. We know \cite{dur'00} that among three-qubit pure states, there exist two inequivalent classes of states, the GHZ- and the W-class which cannot be transformed to each other by stochastic local operations and classical communication (SLOCC) and can be detected by using tangle. Specifically, \(\delta_{\mathcal{C}^2} =0 \) for all the states belonging to the W-class while it is positive for the GHZ-class states. 
We find that after the first round,  $\delta_{\mathcal{C} }^2 $ vanishes for the resulting states, thereby confirming that the three-party output states belong to the W-class. Notice that instead of a weak measurement in Eq. (\ref{eq:weakM}),  if we take a different weak measurement of rank two (mixtures of two Bell states),  we can create GHZ class states too \cite{Pritamnew}. 

\subsection{Non-maximally entangled states are  better  than maximally entangled ones}

Let us  start the protocol with non-maximally entangled state (NME) as resource, given by
\begin{equation}
\label{nent}
  \ket{NME}= \cos z \ket{00}+\sin z \ket{11} , \hspace{0.5 cm} 0\leq z \leq \pi/4.
\end{equation} 
By employing similar procedure as shown for the maximally entangled state, we can again obtain the recursion relation for the output state after an arbitrary round of measurement, say \(n\), in which $\ket{\Psi_{k}^{n}}=\frac{1}{2^{2n-1/2} \sqrt{p_k^n}}\ket{R^n}$ and \(\ket{a^n}\), \(\ket{b^n}\), \(\ket{c^n}\) and \(\ket{d^n}\) get modified accordingly. For example, the output state with the measurement outcome  being \(\sqrt{M_1}\) can be represented as 
 \begin{eqnarray}
 \ket{\Psi_{1}^1}&=&\frac{1}{\sqrt{p_1^1}}\frac{1}{2\sqrt{2}}\bigg[\ket{a^1}\ket{\psi^+}+\ket{b^1}\ket{\psi^-}+\ket{c^1}\ket{\phi^+}\nonumber\\
 &+&\ket{d^1}\ket{\phi^-}\bigg],
 \label{eq:nmes}
\end{eqnarray} 
 where
 \begin{eqnarray}
 \ket{a^1}&=&\sqrt{1+3\lambda}\bigg[\beta_1 \cos z \ket{0} + \alpha_1 \sin z\ket{1}\bigg],\nonumber\\
 \ket{b^1}&=&\sqrt{1-\lambda}\bigg[\beta_1 \cos z\ket{0} - \alpha_1 \sin z\ket{1}\bigg],\nonumber\\
 \ket{c^1}&=&\sqrt{1-\lambda}\bigg[\alpha_1 \cos z\ket{0} + \beta_1 \sin z\ket{1}\bigg],\nonumber\\
 \ket{d^1}&=&\sqrt{1-\lambda}\bigg[\alpha_1 \cos z\ket{0} - \beta_1 \sin z\ket{1}\bigg],\\
 \text{and } \nonumber
 p_1^1 &=& \frac{1}{4}\bigg[1-\lambda \cos 2z \cos \theta \bigg]. 
 \end{eqnarray} 
 Notice that other outcomes of the measurement lead to the same value of GGM after maximizing over state parameters. \\
 
\begin{figure}[ht]
	\includegraphics[width=0.9\linewidth]{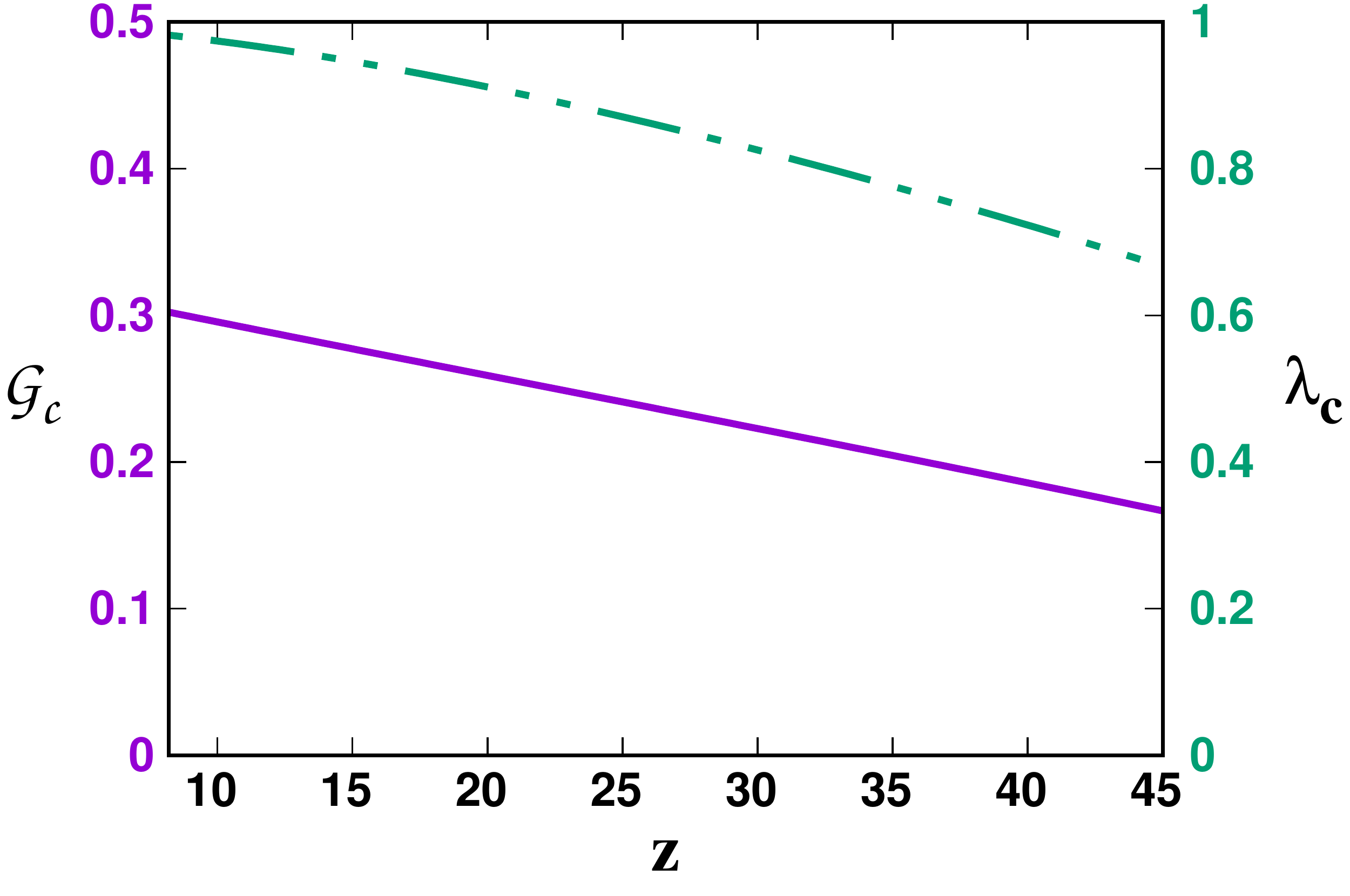}
	\caption{ Critical GGM, $\mathcal{G}_c$ (solid line) (left ordinate) and $\lambda_c$ (solid-dashed line) (right ordinate) against resource state parameter $z$ of \(|NME\rangle\) state in Eq. (\ref{nent}) after the first round. Non-maximally entangled states are better than that of the maximally entangled ones.  The horizontal axis is in degree while the vertical axes are dimensionless. 
	}
	\label{ggm_c}
\end{figure}

 \textbf{Proposition.} \emph{The critical GGM value for the non-maximally entangled state in the first round  of the inflation procedure is higher than that for the maximally entangled state.} \\
\emph{Proof.} As shown for the maximally entangled state, for \(\lambda < \lambda_c\), the maximum eigenvalue contributed for GGM is  given by
\begin{eqnarray}
e_{A_3} &=& \frac{1}{2}+ \frac{\sqrt{X^{A_3}}}{4(1-\lambda\cos2z \cos \theta_1 )}
\end{eqnarray}
with \(X^{A_3} = (1-\lambda)((1+\lambda)(3+\cos 4z) +\sqrt{(1-\lambda)(1+3\lambda)}(1-\cos 4z)-8\lambda\cos 2z \cos \theta_1)\) 
while for \(\lambda > \lambda_c\), it is
\begin{eqnarray}
e_{A_1} &=& \frac{1}{2}+\frac{\sqrt{X^{A_1}}}{4(1-\lambda \cos 2z \cos \theta_1)}, 
\end{eqnarray}
where \(X^{A_1}=2+3\lambda^2 + (2-\lambda^2)\cos 4z -2\lambda\cos 2z (4\cos \theta_1 -\lambda \cos 2z \cos 2\theta_1)\).
We first notice that $\mathcal{G}$ is  independent of \(\phi_1\) of the auxiliary system and is maximized at $\theta_1 = 0$ for any values of $z$.
At $\lambda=\lambda_c$, we have
\begin{eqnarray}
\label{lc}
\nonumber e_{A_3} &=& e_{A_1}\\
\text{or, } \lambda_c &=& \frac{8\left(\cos^4{z}+\csc^2{z}\sqrt{\cos^2{z} \sin^{10}{z}}\right)}{7+\cos{4z}},
\end{eqnarray}
and the corresponding 
\begin{equation}
\label{gc}
   \mathcal{G}_c = 1-[e_{A_1}]_{\lambda=\lambda_c}
\end{equation}
with respect to  $z$ is shown in Fig. \ref{ggm_c} which clearly indicates that non-maximally entangled states outperform the maximally entangled ones  for $z\geq 8.02$.  

\hfill $\blacksquare$
 
Note here that  we have excluded the region $0\leq z <8.02$, since optimizing \(\mathcal{G}\) with respect to \(\lambda\) goes beyond the numerical precision. However, it is clear that the monotonic decrease of \(\mathcal{G}_c\) with the increase of \(z\) has a reverse behavior for small values of \(z\).  

\begin{figure}[ht]
	\includegraphics[width=0.9\linewidth]{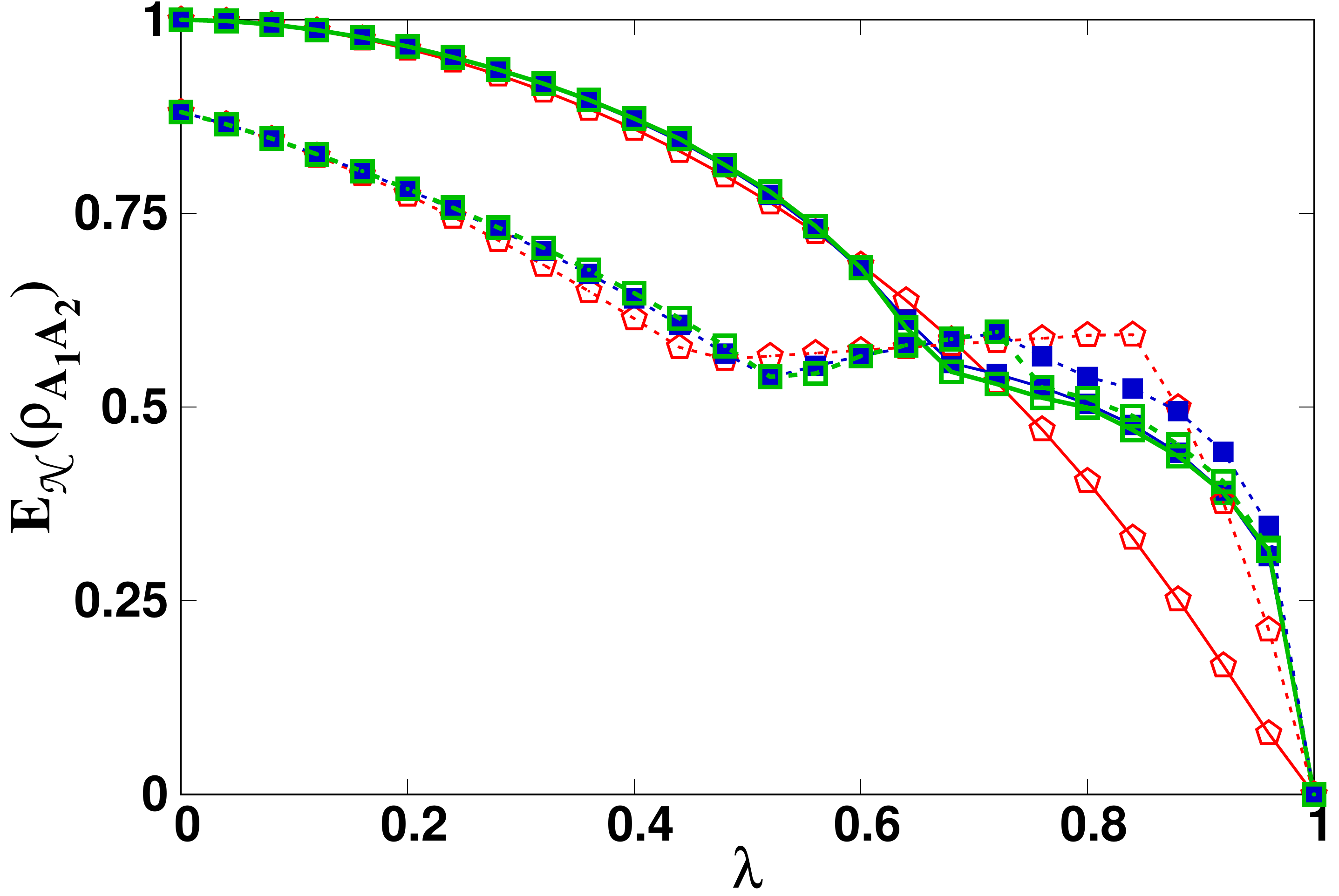}
	\caption{$E_{\mathcal{N}}(\rho_{A_1 A_2} (\lambda))$ (vertical axis)  against $\lambda$ (horizontal axis)  for three-(pentagons), four-(filled squares) and five-party (hollow squares) entangled states produced after measurements. Here, the initial resources are (a) maximally entangled state (solid line),  and (b) NME state with $z=0.5$ (dotted lines) in radian. Both the axes are dimensionless.}
	\label{log_neg}
\end{figure}

Suppose, we want to investigate the  behavior of entanglement of the  reduced density matrix,  $\rho_{A_1 A_2}$, after tracing out the rest of the parties of the output state by  varying the tuning parameter $\lambda$ in  measurement. In this picture, since entangling measurement is performed by a single observer and after measurement, we trace out the ancillary system, it can be represented as  local operations. 
Specifically, we compute logarithmic negativity, $E_{\mathcal{N}}(\rho_{A_1 A_2}(\lambda))$ \cite{zyckohoro'98, zyczkowski'99, vidal'02, plenio'05} in different rounds after optimizing GGM of the output states over auxiliary state parameters. The initial state is NME as well as maximally entangled state as shown in Fig. \ref{log_neg}. As expected, it starts decreasing due to the operation, although the interesting part is the nonmonotonic nature of $E_{\mathcal{N}}(\rho_{A_1 A_2} (\lambda))$ with respect to \(\lambda\). 

\subsubsection{Role of sharpness parameter on resource}

\begin{figure}[ht]
	\includegraphics[width=0.9\linewidth]{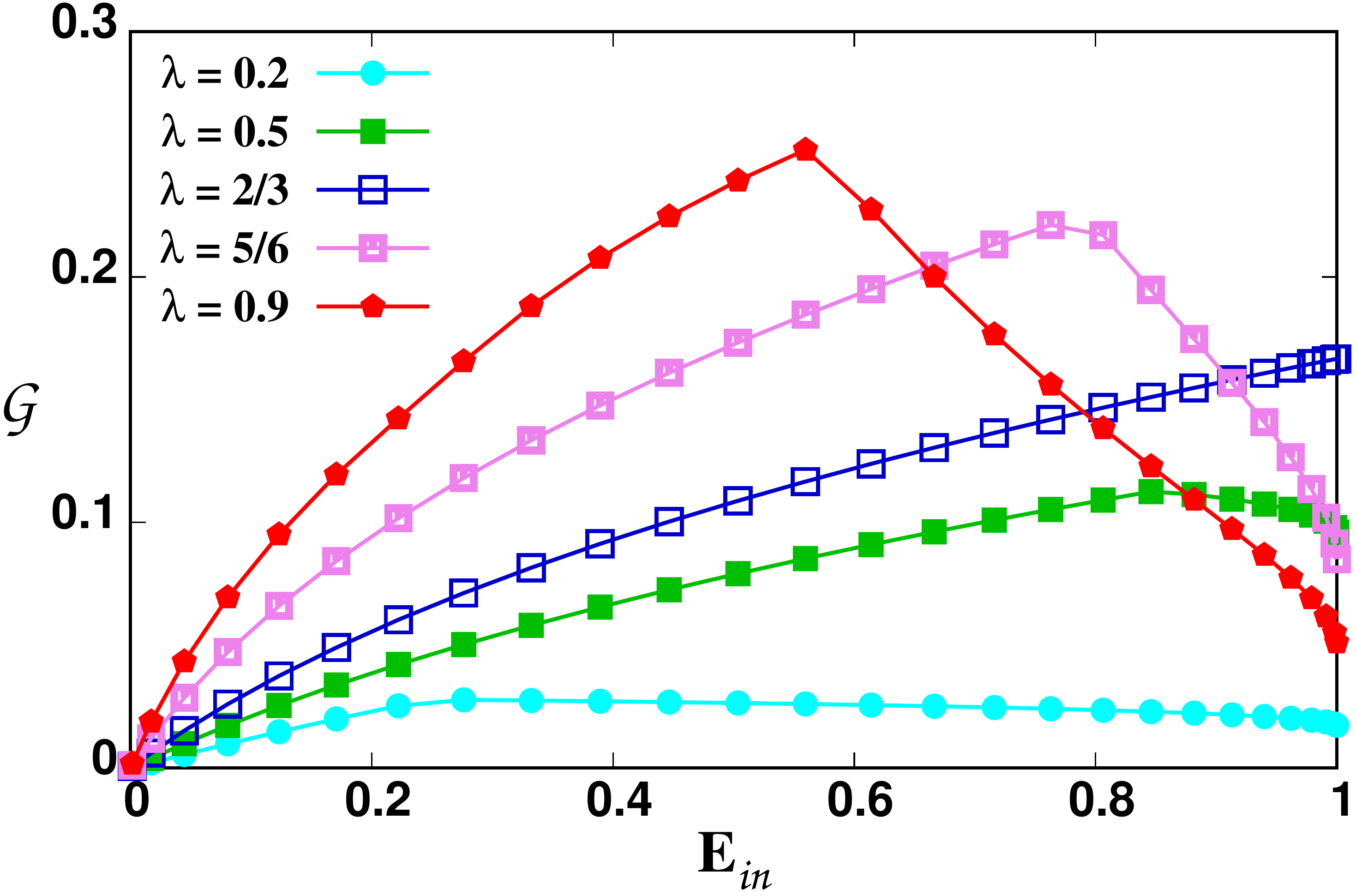}
	\caption{\emph{Non-maximally entangled pure states are better than maximally entangled ones except  for \(\lambda =2/3\).} $\mathcal{G}$ (ordinate) against initial entanglement, $E_{in}$ (abscissa) of the  resource state, i.e. \(\ket{NME}\) for different fixed values of \(\lambda\). Here \(E_{in}\) is the von-Neumann entropy \cite{vn} of the local density matrix of \(|NME_{A_1 A_2}\rangle\).   It shows that maximally entangled  state is best, when $\lambda=\frac{2}{3}$ and for each \(\lambda\) value, there is a unique non-maximally entangled state which can create a maximum genuine multipartite entanglement. The vertical axis is dimensionless while the horizontal axis is in ebits. }
	\label{ient}
\end{figure}

 After optimizing over auxiliary state parameters, the  GGM values of the multipartite state depends drastically on the sharpness parameter \(\lambda\) of the measurement as seen in Fig. \ref{Fig:3}. Analyzing the first round, it is clear that for a fixed \(\lambda\) value, there exists a specific entangled state which can be used as the initial state in the protocol, leading to a maximum multipartite entanglement. For example, when \(\lambda = 2/3\), maximally entangled state is the best resource. Such a relation continues in any round of the protocol. Except for a very small values of \(\lambda\) where almost all the  entangled pure states result to a similar amount of GGM after measurement, for a given \(\lambda\), we find that the moderate amount of entanglement in the resource state is enough to create GME state having maximal GGM as depicted   in Fig. \ref{ient}. It also indicates that the difference between the bipartite entanglement content in the NME state and the maximally entangled one can be compensated by introducing entangling weak measurements like in Eq. (\ref{eq:weakM}) in the production of multipartite entanglement.

\subsubsection{Patterns of genuine multipartite entanglement with increasing round}

\begin{figure}[ht]
	\includegraphics[width=0.9\linewidth]{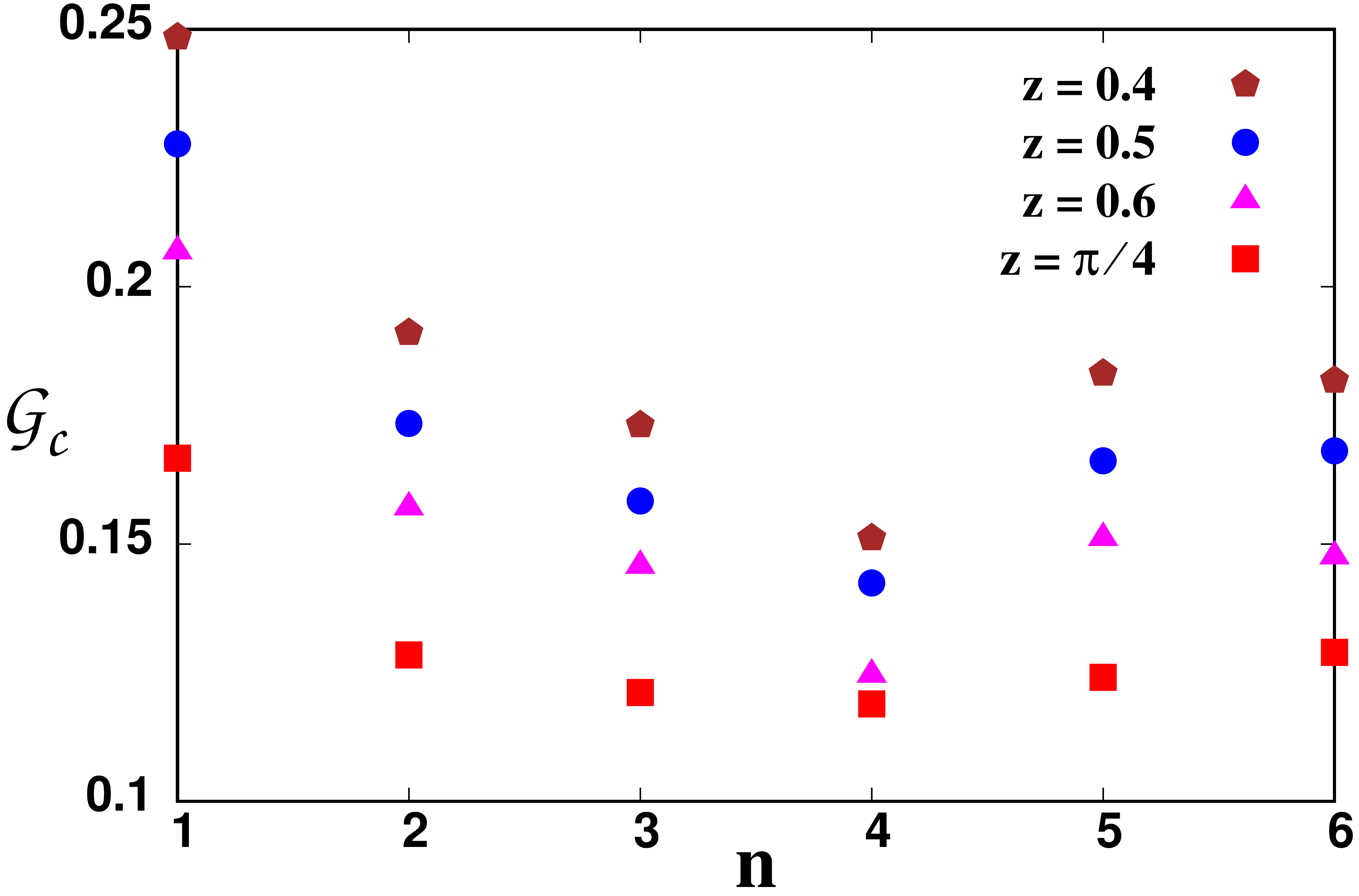}
	\caption{Critical GGM, $\mathcal{G}_c$  (\(y\)-axis) with respect to number of rounds,  $n$ (\(x\)-axis) for different values of $z$ of the NME state. As the number of round increases, thereby the increase of the number of parties in the output state, our optimization algorithm over state parameters becomes inefficient and give a more fluctuating values of \(\mathcal{G}_c\) (up to $10^{-2}$ order) although the  nature of the plot remains almost same. Both the axes are dimensionless. }
	\label{ggmvsround}
\end{figure}

For a maximally entangled state, we observe that the amount of multipartite entanglement decreases with the increase of number of rounds (upto three) (see Fig. \ref{Fig:3}). At this point, there are two natural questions that can be raised --  decreasing trends of GGM  with more number of rounds and  independent  patterns of critical  GGM  for all values of \(z\).  After optimizing both \(\lambda\)  and initial state parameters, we find that even in the case of maximally entangled state, \(\mathcal{G}_c\) decreases with \(n\),  then fluctuate and saturates. For other values of \(z\), the decreasing trend is quite nicely visible from Fig. \ref{ggmvsround} for small \(n\), and then the fluctuation is also quite high for other values of \(z\).  
%

Note also that, with the  increasing round, number of parameters over which we have to maximize also increases. In the $n$th round, we need to maximize over $2n$ auxiliary system variables and also over $\lambda$, i.e., total $2n+1$ variables, thereby making  the optimization problem harder with each round. Moreover, with the increase of number of parties, calculating GGM becomes also complicated as one has to consider all the bipartitions. Up to fifth rounds, we have computed all the bipartition for computing GGM, and find that the  contribution in $\mathcal{G}_c$ is  coming from eigenvalues of the reduced density matrices, $\rho_{A_1} \equiv \rho_{A_{n+2}} \equiv \rho_{A_{n+1} A_{n+2}}$. And hence unless mentioned otherwise, for the sixth round onwards, GGM is computed by taking all  single-  and nearest neighbor  two-site density matrices.

\subsection{Haar-uniformly generated two-qubit pure states as initial state for inflation}
\label{rand}

\begin{figure}[ht]
	\includegraphics[width=0.9\linewidth]{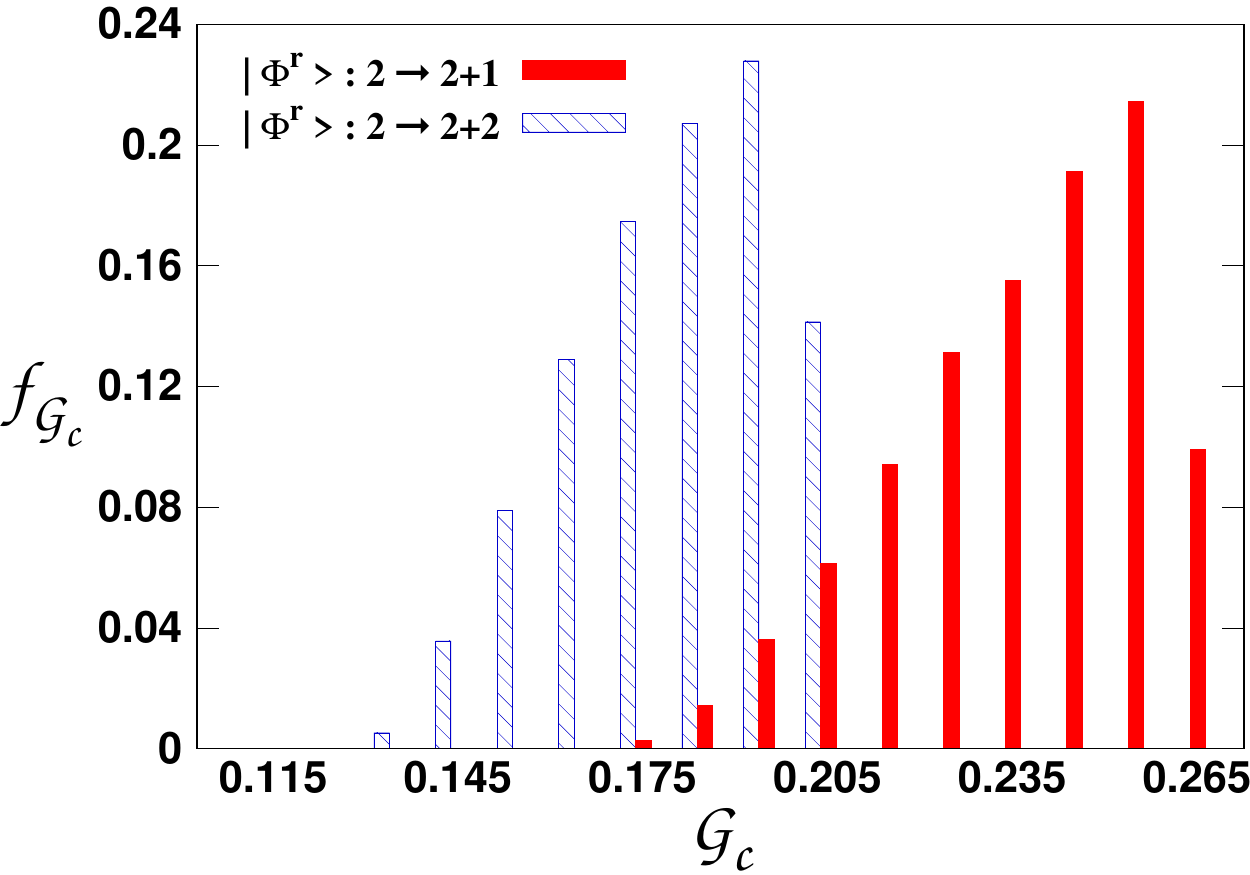}
	\caption{Normalized frequency distribution, $f_{\mathcal{G}_c}$, (vertical axis), against $\mathcal{G}_c$ (horizontal axis). We Haar uniformly generate random pure states and create three- (solid bars) and four-party (check bars) GME states after maximization over state and measurement parameters. In each case, the sample size is taken to be $5 \times 10^3$. Both the axes are dimensionless. }
	\label{Fig:5}
\end{figure}

Instead of a  specific class of pure states, let us find out the universal trends  of GGM for the resulting state when the initial resource state is chosen Haar uniformly \cite{bengtsson'06}, given by
 $|\phi^{r}\rangle = a\ket{00}+b\ket{01}+c\ket{10}+d\ket{11}$ with  \(a = a' + i a''\),  similarly \(b\), \(c\), \(d\),  and \(a', a'',\ldots\)   being chosen randomly from Gaussian distribution with mean \(0\) and standard deviation unity. Opting similar technique like the maximally entangled state, we can again write down the output state after round \(n\) as 
 \begin{equation*}
    \ket{\Psi_{k}^{n}}=\frac{1}{2^{2n-1/2} \sqrt{p_k^n}}\ket{R_n},
\end{equation*}
 where \(\ket{R_n} \) can be suitably obtained via recursion relation in terms of \(\ket{a^n}\), \(\ket{b^n}\), \(\ket{c^n}\), and \(\ket{d^n}\)  as in Eq. (\ref{eq:abcd}). Notice that the change of resource states only effects the form of  \(\ket{a^1}\), \(\ket{b^1}\), \(\ket{c^1}\), and \(\ket{d^1}\).    For example, after the first round with \(\sqrt{M_1}\) being the outcome of the measurement, the expression for the resulting state is similar to the one given in Eq. (\ref{eq:nmes}) with 
 \begin{eqnarray*}
 &&\ket{a^1}=\sqrt{1+3\lambda}\bigg[\beta_1\big(a\ket{0}+c\ket{1}\big)+\alpha_1\big(b\ket{0}+d\ket{1}\big)\bigg],\\&&
 \ket{b^1}=\sqrt{1-\lambda}\bigg[\beta_1\big(a\ket{0}+c\ket{1}\big)-\alpha_1\big(b\ket{0}+d\ket{1}\big)\bigg],\\&&
 \ket{c^1}=\sqrt{1-\lambda}\bigg[\alpha_1\big(a\ket{0}+c\ket{1}\big)+\beta_1\big(b\ket{0}+d\ket{1}\big)\bigg],\\&&
 \ket{d^1}=\sqrt{1-\lambda}\bigg[\alpha_1\big(a\ket{0}+c\ket{1}\big)-\beta_1\big(b\ket{0}+d\ket{1}\big)\bigg]. 
 \end{eqnarray*} 
Let us study the behavior of critical GGM,  \(\mathcal{G}_c\), for Haar uniformly generated pure states in details after the first and the second rounds. Towards this, we calculate the normalized frequency distribution of   \(\mathcal{G}_c\), denoted by  \(f_{\mathcal{G}_c} = \frac{N(\mathcal{G}_c)}{N}\), where \(N(\mathcal{G}_c)\) is the number of Haar uniformly generated state having a fixed \(\mathcal{G}_c\) value and \(N\) is the total number of states simulated.  
The frequency distribution, given in Fig. \ref{Fig:5}, indicates the following observations which are in good agreement with the previous results. 
\begin{itemize}
\item We again observe that  at \(\lambda_c\),  \(\mathcal{G}_c \)  obtained from non-maximally entangled random states  is  higher than that of maximally entangled states.  We know that the average entanglement in the random two-qubit pure states is around \(0.48\) \cite{rivu'20} and as discussed before, if the initial state contains a certain entanglement value, it is always possible to tune the sharpness parameter in such a way that  the resulting state has more genuine multipartite entanglement than that of the maximally entangled state. It again establishes that  a trade-off relation between sharpness parameter involved in the entangling measurement and the entanglement content of the resource state plays an important role towards the success of the protocol.

\item After second round of the protocol, the mean  of \(\mathcal{G}_c = 0.182\) decreases compared to the first round which is \(0.237\), as shown in Fig. \ref{Fig:5}.  Similarly, the standard deviations of the distribution in the first and the second rounds are respectively \(0.02\) and \(0.017\). 


\item All the resulting tripartite state after the first round belong to the W-class. 
\end{itemize}

\subsection{Creation of multipartite entangled state from noisy entangled two-qubit initial state }

We move to a more realistic situation where the initial shared state is noisy. In particular, we take the Werner state \cite{werner'89},
\begin{eqnarray}
\label{werner}
\rho^W = p\ket{\phi^+}\bra{\phi^+}+(1-p)\frac{\mathbb{I}_4}{4}
\end{eqnarray}
as the initial resource. One of the main obstacle in this situation is to quantify multipartite entanglement content of the output state after each round. To overcome it, we compute negativity monogamy score \cite{coffman'00, dhar'17} which measures the distribution of entanglement in a multipartite state. Recently, it was also argued that the overall behavior of monogamy scores is quite similar to the multipartite entanglement measures \cite{soorya'19}.

Taking initial state as $\rho^W\otimes \rho_{A_3} \equiv \rho^W \otimes \ket{\chi_1^+}\bra{\chi_1^+}$, and performing  weak measurement on the part of the Werner state and the auxiliary system, we compute \(\delta_{\mathcal{N}_c} = \max \delta_{\mathcal{N}}\) where maximization is performed over the  coefficients \(\theta_1\), \(\phi_1\) of the auxiliary state and \(\lambda\) of the weak measurement.  In the second round, we take the resulting state of the first round as the initial state and another auxiliary state, i.e., \( \rho_{A_1 A_2 A_3} \otimes \ket{\chi_2^+}\bra{\chi_2^+}$ and so on. 
In the first round, by obtaining the outcome $\sqrt{M_1}$, the tripartite state reads
\(\rho_1^1=\frac{\sqrt{M_1}\left(\rho^W\otimes \rho_{A_3}\right)\sqrt{M_1}}{p_1^1}\), 
where 
\(p_1^1 = \mbox{Tr}\left(\sqrt{M_1}\left(\rho^W\otimes \rho_{A_3}\right)\sqrt{M_1}\right) \). 

\begin{figure}[ht]
	\includegraphics[width=0.9\linewidth]{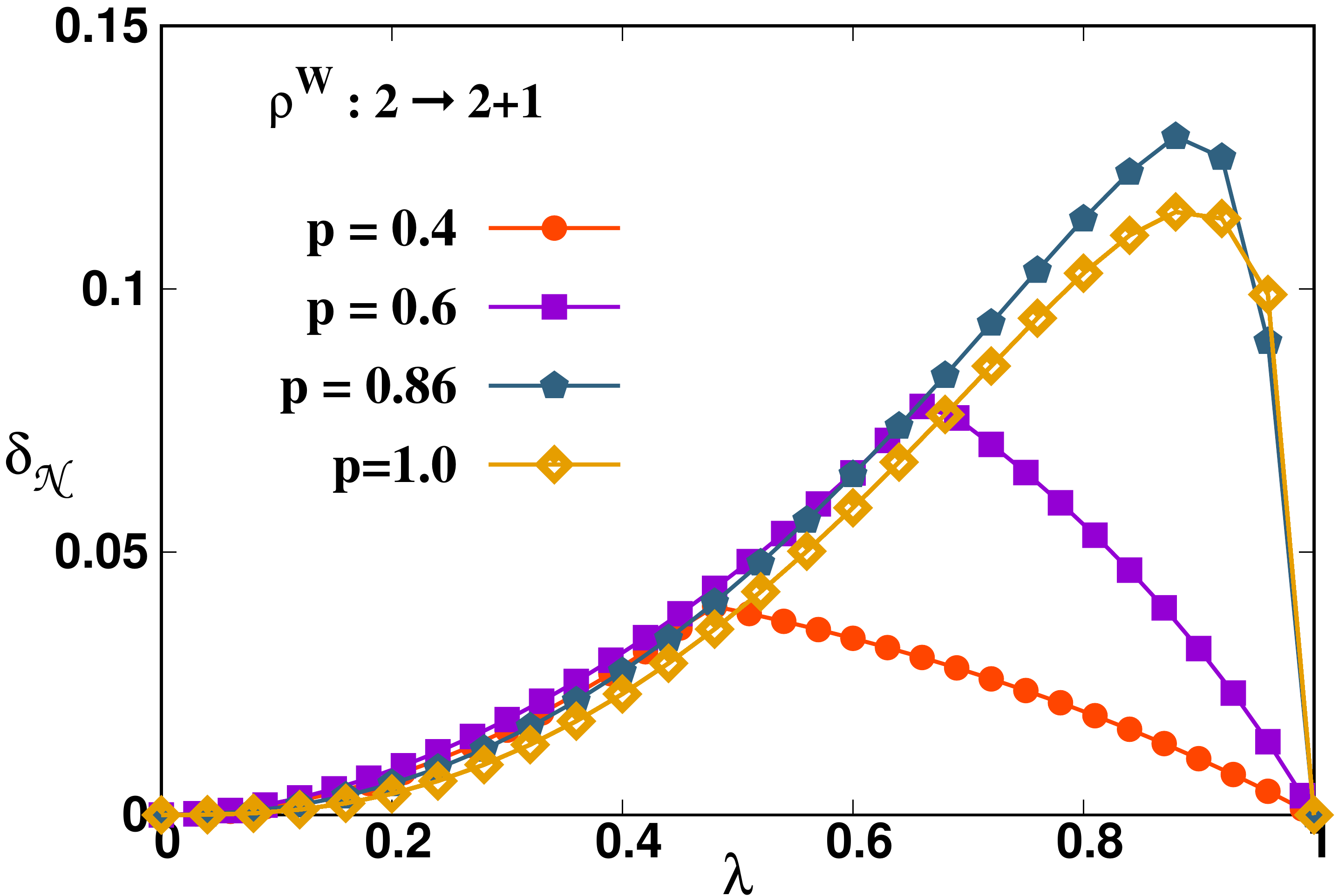}
	\caption{ Optimized negativity monogamy score, $\delta_{\mathcal{N}}$ (vertical axis) with the sharpness parameter $\lambda$ (horizontal axis), for different values of noise parameter, $p$ in the Werner state, $\rho^{W}$. Both the axes are dimensionless.}
	\label{Fig:6}
\end{figure}

After maximizing \(\delta_{\mathcal{N}}\) with respect to variables $\theta_1, \phi_1$ of auxiliary system,  the variation of negativity monogamy score of $\rho_1^1$ with \(\lambda\)  for different values of the noise parameter $p$ is shown in Fig. \ref{Fig:6}. Like pure states, we also observe here that for a fixed value of \(p\), there is a unique critical \(\lambda\) value up to which  \(\delta_{\mathcal{N}}\) increases with \(\lambda\) and then starts decreasing with \(\lambda\).   Notice here that the probability of obtaining any of the four outcomes $\{M_k\}_{k=1}^4$ is again $p_k^1=\frac{1}{4}$ in the first round. 

\begin{figure}[ht]
	\includegraphics[width=0.9\linewidth]{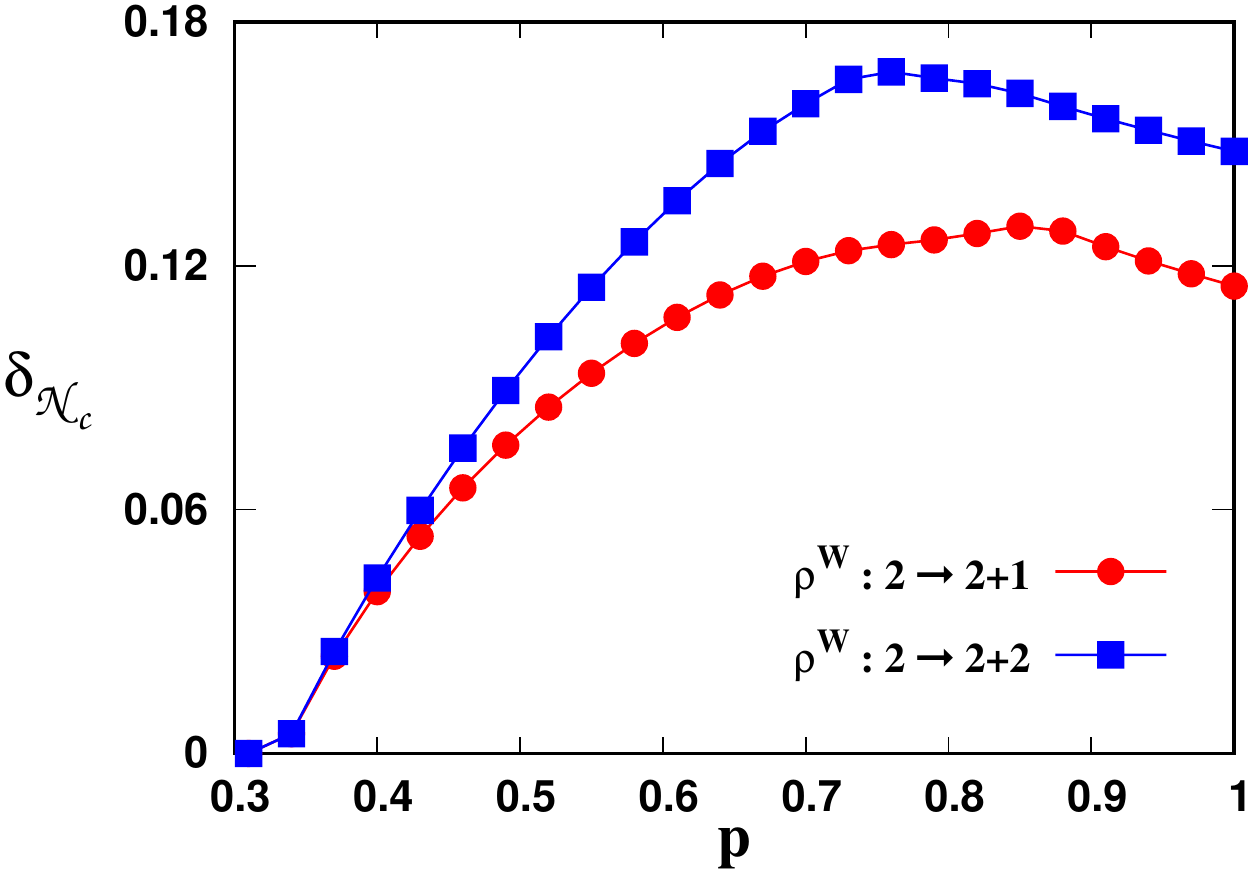}
	\caption{Critical negativity monogamy score, $\delta_{\mathcal{N}_c}$ (ordinate), against noise parameter $p$ (abscissa) of the initial resource $\rho^W$ (Werner state) after the first (circles) and the second rounds (squares), i.e. in the inflation process, \(2 \rightarrow 2+1, 2+2\). The maximum $\delta_{\mathcal{N}_c}$ obtained for \(p \neq 1\) (i.e., other than unity) ensures that noisy entangled states can create  a high amount of multipartite entanglement than that of its noiseless counterparts.   Both the axes are dimensionless. }
	\label{max_mon}
\end{figure}

Interestingly, the negativity monogamy score does not behave monotonically with the increase of  noise, \(p\) in the resource state. To visualize it, we consider  the behavior of the critical monogamy score, \(\delta_{\mathcal{N}_c}\) with the increase of \(p\) (see Fig. \ref{Fig:6}). At each round,  we observe that the negativity monogamy score reaches its maximum value when the state is noisy compared to the pure state which again demonstrates that there is a competition between the  entanglement in the measurement and in the resource state. For example, in the first round, the maximum of \(\delta_{\mathcal{N}_c}\)  occurs at \(p = 0.858\) while it is \(p= 0.75\) after completion of the second round, thereby showing  robustness in monogamy of entanglement against noise. Notice also that unlike GGM, \(\delta_{\mathcal{N}_c}\) increases with the number of rounds, as shown in Fig. \ref{max_mon}.. 

\section{Multiqubit  Pure States for Expanding Multipartite Entangled State (\(3 \rightarrow 3 +n\)): W state is better than GHZ state}
\label{sec:multitomulti}

Let us now change the resource state from a bipartite state to  a tripartite one by keeping the product auxiliary systems and  the same unsharp measurements as in Eq. (\ref{eq:weakM}). We generate  three-qubit states Haar uniformly both from the GHZ- as well as the W-class and compare their potential to expand genuine multipartite entanglement in higher number of qubits via the weak measurement strategy.  Before that, let us consider two important class of  tripartite states, namely the GHZ 
state, given by 
$\ket{GHZ}=\frac{1}{\sqrt{2}}\big(\ket{000}+\ket{111}\big)$  and the W state,  \(\ket{W}=\frac{1}{\sqrt{3}}\left(\ket{001}+\ket{010}+\ket{100}\right)\) as inputs. 

Interestingly,  by using the similar recursion relation derived for the maximally entangled state, we can show that  starting from \( \ket{GHZ}\otimes \Pi_{i=1}^{n} \ket{\chi_i^+}\), the maximal GGM obtained after maximizing over auxiliary states  is equal to the GGM of the initial state having maximally entangled state as resource in the \(n\)th round, i.e., for the initial state  \(\ket{\phi^+}\otimes \Pi_{i=1}^{n} \ket{\chi_i^+} \) after measurements for each value of \(\lambda\). Such a correspondence also holds between the generalized GHZ state,   \(\ket{gGHZ}= \cos{z} \ket{000}+\sin{z}\ket{111}\) and the non-maximally entangled  two-qubit  states, \(\ket{NME}\), i.e., 
\(\mathcal{G}_c (\ket{gGHZ}) = \mathcal{G}_c (\ket{NME}) \) for a fixed value of \(z\).   

%
%
In case of the W state, the output state after the first measurement  with the outcome \(\sqrt{M_1}\) reads as
\begin{eqnarray}
\ket{\Psi_{1}^1}&=&\frac{1}{\sqrt{p_1^1}}\frac{1}{2\sqrt{6}}\bigg[\ket{a^1}\ket{\psi^+}+ \ket{b^1}\ket{\psi^-}+\ket{c^1}\ket{\phi^+} \nonumber\\
&+&\ket{d^1}{\ket{\phi^-}}\bigg],
\label{eq:woutput}
\end{eqnarray}
where \(p_1^1\) is the probability of obtaining \(\sqrt{M_1}\), and  
\begin{eqnarray}
&&\ket{a^1}=\sqrt{1+3\lambda}\big(\sqrt{2}\beta_1\ket{\psi^+}+\alpha_1\ket{00}\big),\nonumber\\
&&\ket{b^1}=\sqrt{1-\lambda}\big(\sqrt{2}\beta_1\ket{\psi^+}-\alpha_1\ket{00}\big),\nonumber\\
&&\ket{c^1}=\sqrt{1-\lambda}\big(\sqrt{2}\alpha_1\ket{\psi^+}+\beta_1\ket{00}\big),\nonumber\\
&&\ket{d^1}=\sqrt{1-\lambda}\big(\sqrt{2}\alpha_1\ket{\psi^+}-\beta_1\ket{00}\big).
\label{eq:Wcoeff}
\end{eqnarray}
\begin{figure}[ht]
	\includegraphics[width=0.9\linewidth]{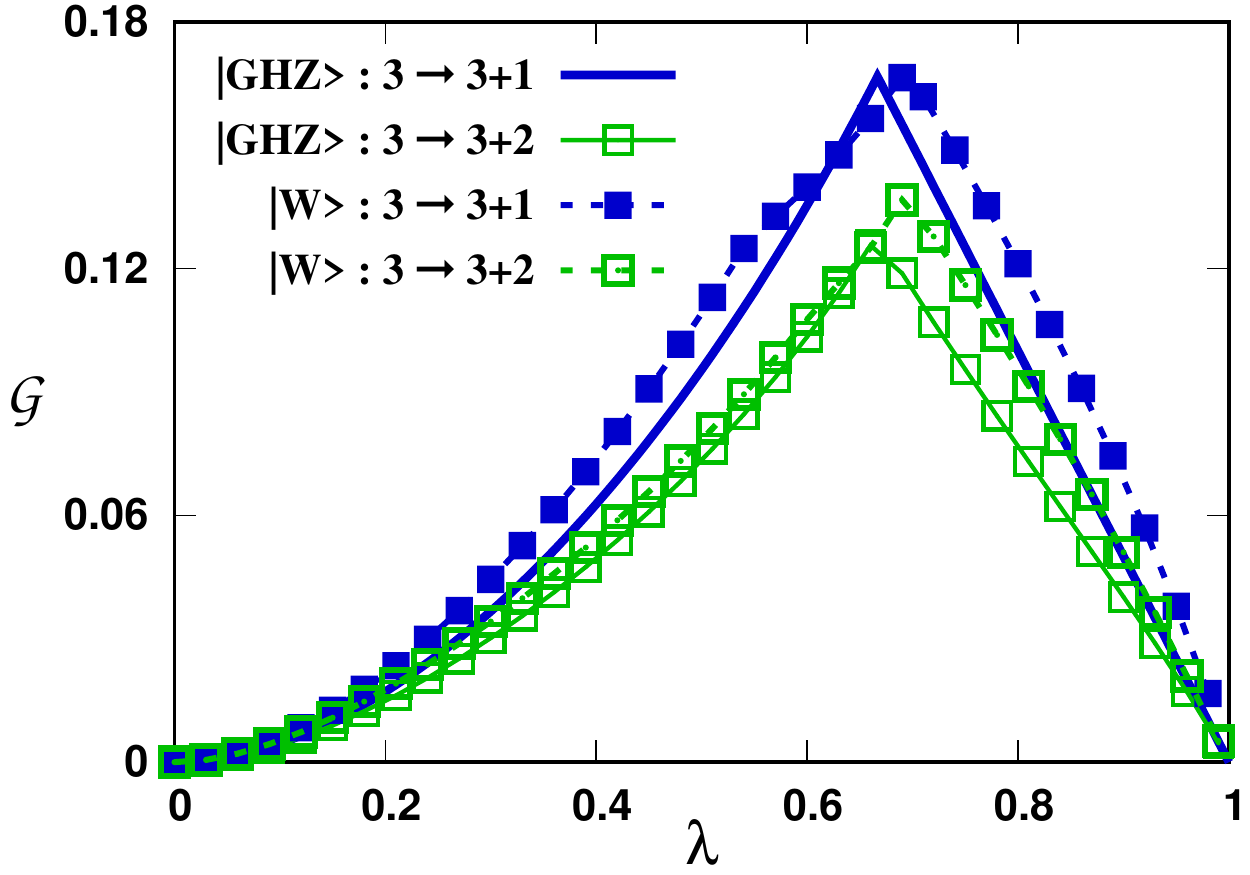}
	\caption{\emph{GHZ vs. W states.} Maximized $\mathcal{G}$ (ordinate) vs. $\lambda$ (abscissa). We create four-  (filled squares) and five-party (hollow squares) GME states by taking $\ket{GHZ}$ (solid lines) and  $\ket{W}$  (dashed lines) states as initial resources. In the former case,  $\lambda_c = 2/3$ while the latter case, \(\lambda_c \approx 0.693$. Both the axes are dimensionless.  }
	\label{w_ggmc}
\end{figure}
Before performing optimization over \(\lambda\), we study the behavior of \(\mathcal{G}\) after optimizing over \(\{\theta_1, \theta_2, \phi_1, \phi_2\}\) as seen in Fig. \ref{w_ggmc}. In the first round, for the W state, \(\mathcal{G}_c = 0.168$ while in the second round, it becomes $0.138$ and  $\lambda_c=0.693$ is same for both the rounds. Notice that in case of the shared GHZ state, \(\lambda_c = 2/3 = 0.678\) and the corresponding \(\mathcal{G}_c = 0.167 \) and \(0.128\) in the first and the second rounds respectively. In the literature, the class of GHZ states are typically shown to be more useful than the W-type states although there are counterexamples \cite{Wstate1, Wstate2, Wstate3, Wstate4, Wstate5}.  Our results indicate that entanglement inflation  is another process which can show  benefit of sharing  W state.

\subsubsection{Spreading  entanglement via random three-qubit states: GHZ-class vs. W-class}

\begin{figure}[ht]
	\includegraphics[width=0.9\linewidth]{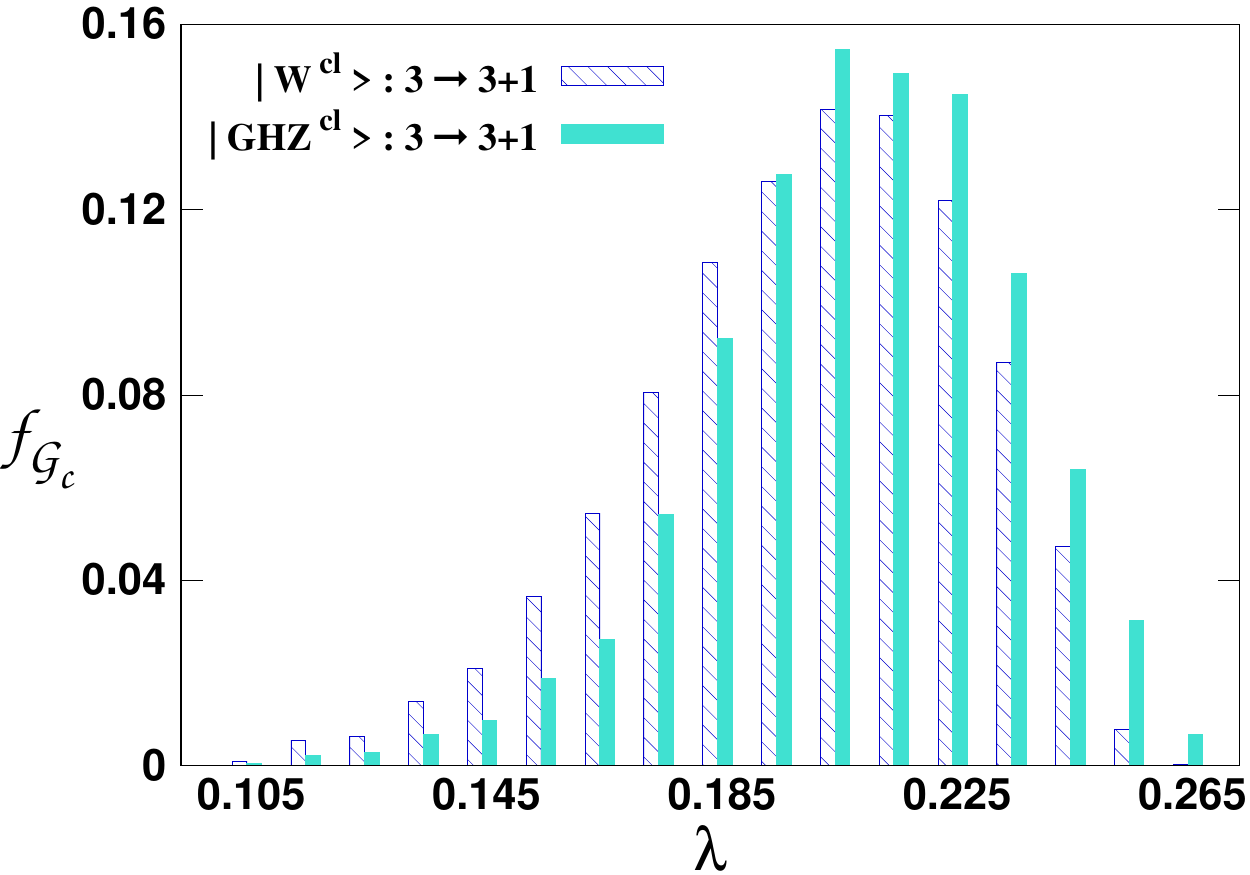}
	\caption{Normalized frequency distribution, $f_{{\mathcal{G}_c}}$ (vertical axis) against critical GGM, $\mathcal{G}_c$ (horizontal axis).  Haar uniformly generated random three-qubit GHZ- (solid bars) and W-class states (check bars) are used to generate four-party state in the \(3\rightarrow 3+1\) inflation method in an optimal way.
	 In each case, the  sample size is $5 \times 10^3$. Both the axes are dimensionless.  }
	\label{ghzwclass_ggm}
\end{figure}

Let us generate three-qubit states Haar uniformly, which belong to the GHZ-class, given by 
\( \ket{GHZ^{cl}}=a\ket{000}+b\ket{010}+c\ket{001}+d\ket{100}+ e\ket{011}+f\ket{101}+g\ket{110}+h\ket{111}\) where the coefficients are complex and are chosen from Gaussian distribution as discussed in case of two-qubit random states. 
After the outcome $\sqrt{M_1}$,  we obtain  the resulting state, \(\ket{\Psi_{1}^{1}}=\frac{1}{\sqrt{p_1^1}} \frac{1}{2\sqrt{2}} \ket{R^1}\) which is in the same form as in Eq. (\ref{eq:woutput}) where the  coefficients can be modified as
\begin{eqnarray*}
&&\ket{a^1} = \sqrt{1+3\lambda}\big(\beta_1\ket{X}+\alpha_1\ket{Y}\big), \\
&&\ket{b^1} = \sqrt{1-\lambda}\big(\beta_1\ket{X}-\alpha_1\ket{Y}\big),\\
&&\ket{c^1} = \sqrt{1-\lambda}\big(\alpha_1\ket{X}+\beta_1\ket{Y}\big), \\
&&\ket{d^1} =\sqrt{1-\lambda}\big(\alpha_1\ket{X}-\beta_1\ket{Y}\big)\\
\text{and }
&&\ket{X} = a\ket{00}+b\ket{01}+d\ket{10}+g\ket{11}, \\
&&\ket{Y} = c\ket{00}+e\ket{01}+f\ket{10}+h\ket{11}. 
\end{eqnarray*}
After \(n\) rounds, the output state corresponding to the outcome, \(\sqrt{M_k}\),  becomes  \( \ket{\Psi_{k}^{n}} = \frac{1}{\sqrt{p_k^n}}\frac{1}{2^{2n-1/2}} \ket{R^n}\) where \(\ket{a^n}, \ket{b^n}, \ket{c^n}\) and \(\ket{d^n}\) can be written in terms of  \(\ket{a^1}, \ket{b^1}, \ket{c^1}\) and \(\ket{d^1}\) . 
The normalized frequency distribution  of $\mathcal{G}_c$ after  the first round is shown in Fig. \ref{ghzwclass_ggm}, having mean \(0.208\) and standard deviation \(0.026\). Comparing this average value with that of  the Haar uniformly generated two-qubit states, we find that two-qubit states in this process is more resourceful than that of the multiqubit states on average.  

Similar analysis can be performed by simulating    W-class states, $|W^{cl}\rangle = a\ket{000}+b\ket{010}+c\ket{001}+d\ket{100}$, Haar uniformly and  optimize the GGM value over \(\lambda\) and a set of parameters \(\{\theta_1, \phi_1\}\) in the auxiliary system. In this case, the resulting state can be represented similarly as in the case of the GHZ-class except the coefficients 
gets modified as
\begin{eqnarray*}
\ket{a^1} &=& \sqrt{1+3\lambda}\big(\beta_1\ket{Z}+\alpha_1 c \ket{00}\big),\\
\ket{b^1} &=& \sqrt{1-\lambda}\big(\beta_1\ket{Z}-\alpha_1 c \ket{00}\big),\\
\ket{c^1} &=& \sqrt{1-\lambda}\big(\alpha_1\ket{Z}+\beta_1 c \ket{00}\big),\\
\ket{d^1} &=& \sqrt{1-\lambda}\big(\alpha_1\ket{Z}-\beta_1 c \ket{00}\big),\\
\text{and} 
\ket{Z} &=& a\ket{00}+b\ket{01}+d\ket{10}.
\end{eqnarray*}	
In this case, the mean, \(\langle\mathcal{G}_c\rangle = 0.20\) and the standard deviation, \(\sigma_{\mathcal{G}_c} = 0.028\)  of the frequency distribution of the critical GGM which is quite close to that obtained for the GHZ-class states. 

\begin{figure}[ht]
	\includegraphics[width=0.9\linewidth]{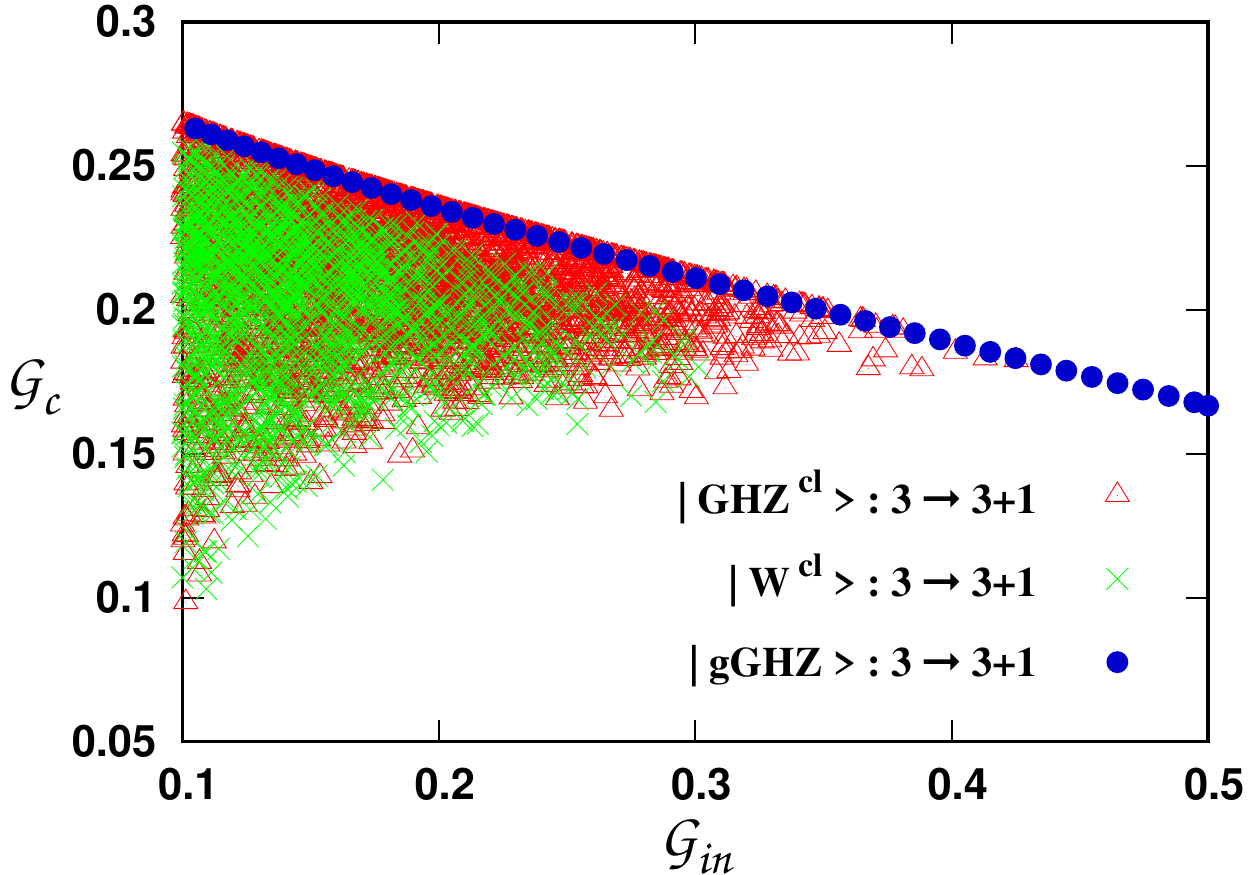}
	\caption{Scattered plot of $\mathcal{G}_c$ (\(y\)-axis) against GGM of the input states, $\mathcal{G}_{in}$ (\(x\)-axis) in \(3\rightarrow 3 +1\) process.  Input states are generalized GHZ (circles) and  Haar uniformly generated GHZ-(triangles), and  W-class (crosses) states. Clearly, producing GME states from the gGHZ state is the best  option compared to the situations when the initial states are  Haar uniformly generated three-qubit states. The sample size in both the GHZ- and W-class states are $5 \times 10^3$. Both the axes are dimensionless. }
	\label{alliggm}
\end{figure}

Let us observe the behavior of \(\mathcal{G}_c\)  against the GGM of the initial shared state which reveals the role of entanglement content of the inputs in this process (see Fg. \ref{alliggm}). We find that for same values of GGM, \(\mathcal{G}_{in}\), in the arbitrary three-qubit state and the generalized GHZ state,    after the first round, the  critical GGM values of the final state obtained from the arbitrary three-qubit states (irrespective of the class) is bounded above by that of the \(\ket{gGHZ}\) state, i.e., \(\mathcal{G}_c (\ket{gGHZ}) \geq \mathcal{G}_{c} (\ket{GHZ^{cl}}) (\mathcal{G}_{c} (\ket{W^{cl}}))\).  It manifests that among all the three-qubit states, the generalized GHZ state is the best resource for expanding genuine multipartite entanglement.  

%
%

\section{ Multiple copies of entangled resource are not always powerful for inflating multipartite entangled states}
\label{sec_pic2}

\begin{figure}[ht]
	\includegraphics[width=0.9\linewidth]{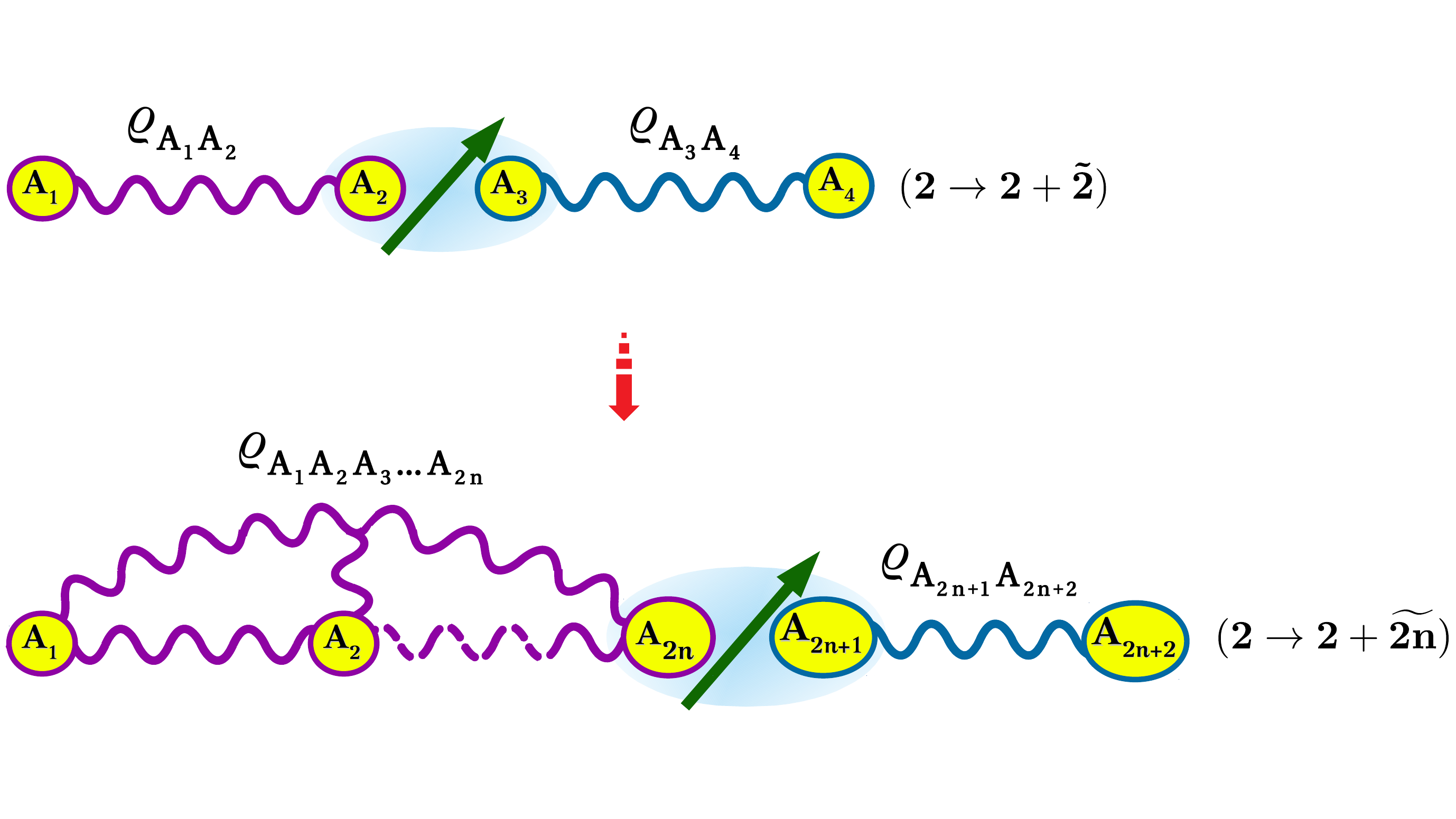}
	\caption{\emph{Entanglement-based inflation scheme.} Schematic diagram of the inflation protocol where multiple copies of the initial resources are available. Unlike the  previous case, the resource state, \(\varrho_{A_1A_2}\) is used as auxiliary system between the node \(A_3\) and \(A_4\). In this situation, \(A_2\) and \(A_3\) perform the weak entangling measurement to create a multipartite entangled state and so on. Importantly, we notice that in this case, the optimization over parameters of the auxiliary system is no more required. This scenario is exactly same as the entanglement swapping performed in a chain. However, instead of weak measurements, if one performs projective measurements in nodes, those nodes gets disconnected, thereby  no production of multipartite entangled state. To discriminate this protocol from the previous one, we denote it as \(2 \rightarrow 2 + \widetilde{2n}\).     }
	\label{Fig:EBchematic}
\end{figure}

Let us change the gear and analyze the situation when the resources are increased. Specifically, in the previous scenario, an entangled state and multiple copies of auxiliary states are initially given while in this scenario, several copies of  entangled states are used as resource as shown in Fig. \ref{Fig:EBchematic}. Specifically, \(n\) copies of the initial state are shared where unsharp measurements (e.g. measurements in  Eq. (\ref{eq:weakM})) are applied in a chain to expand multipartite entanglement -- it is a modified version of entanglement swapping \cite{zukowski'93, swapping2, swapping3, swapping4, swapping5, swapping6} with unsharp measurement.  The major difference of this protocol and the original entanglement swapping protocol is that it transfers entanglement from a pair to an another pair while the protocol via unsharp measurement can create \((2n+2)\)-party state after \(n\) rounds. Since the difference between the previous and current methods is the choice of auxiliary states,  we refer this method as entanglement-based (EB) inflation compared to the previous product-based one. The idea is again to maximize GGM with respect to sharpness parameter, \(\lambda\) in Eq. (\ref{eq:weakM}).

By employing similar technique as discussed in the preceding section, we can again find the recursion relation of the resulting state. Since we want to find the advantages between the previous protocols and the entanglement-based one, we consider again maximally, non-maximally, Haar uniformly simulated entangled states and the Werner state. Notice, however, that the resources used in this protocol grows with \(n\) which in the previous case, remains constant to the initial entanglement. Moreover, it should be mentioned that  in this scenario, one may  invoke
much more general measurement schemes to optimise GME states.

\begin{figure}[ht]
	\includegraphics[width=0.9\linewidth]{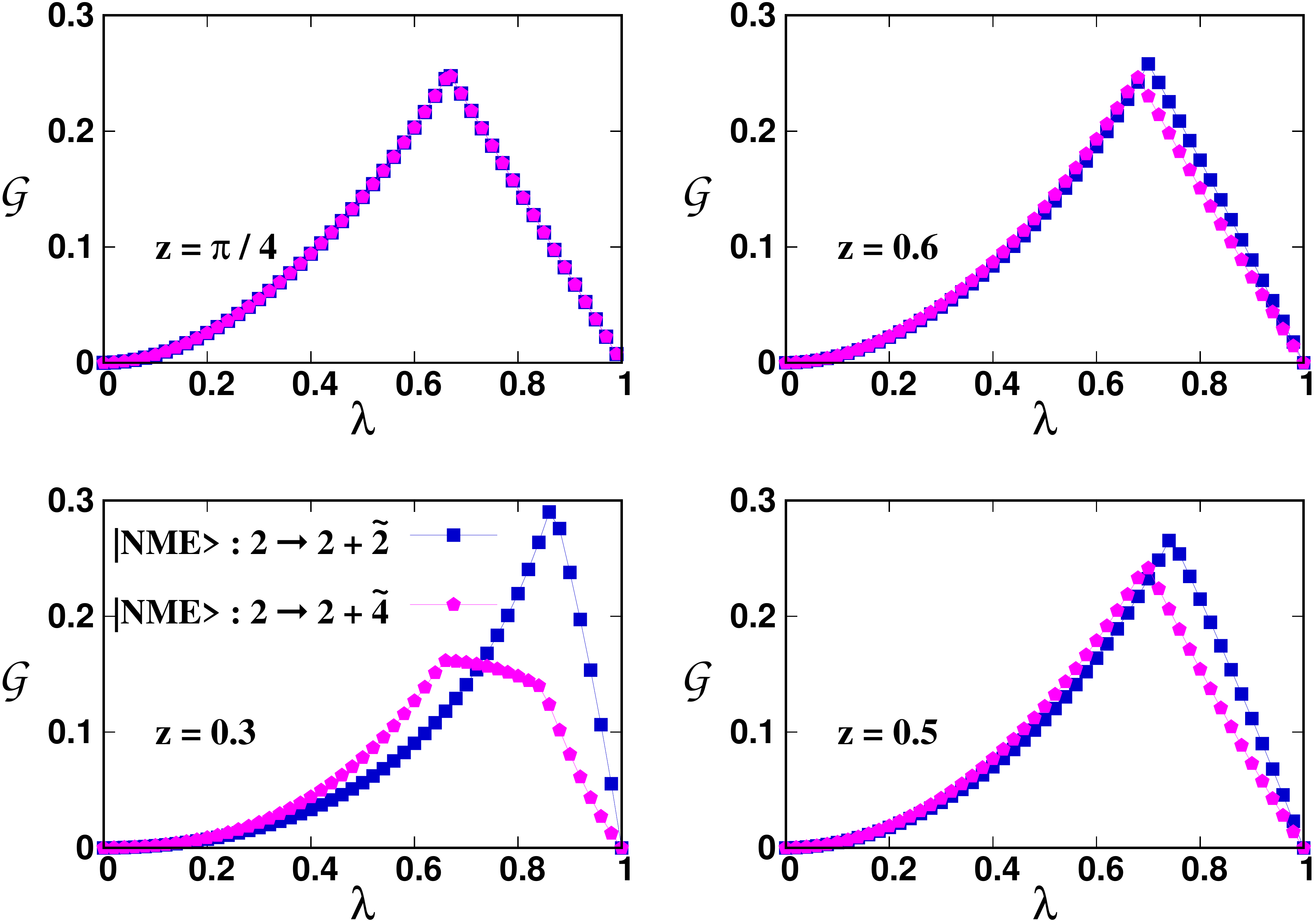}
	\caption{Plot of $\mathcal{G}$ (ordinate) vs. $\lambda$ (abscissa) in the EB inflation process. Four- (squares)  and six-party (pentagons)  GME states are generated by taking different NME states (for different values of  $z$ mentioned in each plots). Both the axes are dimensionless. }
	\label{NMS}
\end{figure}

\emph{Non-maximally entangled pure states.} Before considering \(|NME\rangle\), let us first take $\ket{\phi^+}$ as the initial resource and another $\ket{\phi^+}$ as the auxiliary system i.e. the initial state, $\ket{\phi^+}\otimes\ket{\phi^+}$ and make the joint measurement on the second and  the third party, we obtain the four-party state as
\begin{eqnarray}
\ket{\Psi_{k}^{1}}&=&\frac{1}{2\sqrt{2}}\bigg[\left(\ket{0}\ket{F_+^1} + \ket{1}\ket{E_-^1}\right)\ket{0} \nonumber\\
 &+& \left(\ket{0}\ket{E_+^1} + \ket{1}\ket{F_-^1}\right)\ket{1}\bigg] \nonumber \\ 
 &=& \frac{1}{2\sqrt{2}} \bigg[\ket{Y^1}\ket{0} + \ket{Z^1}\ket{1}\bigg], 
\label{eq:EBmax}
\end{eqnarray}
while after the second round, i.e., after  measuring the fourth and the fifth party, the six-party state becomes
\begin{eqnarray*}
\ket{\Psi_{k}^{2}}&=&\frac{1}{4\sqrt{2}}\bigg[\left(\ket{Y^1}\ket{F_+^2} + \ket{Z^1}\ket{E_-^2}\right)\ket{0}\\ &+& \left(\ket{Y^1}\ket{E_+^2} + \ket{Z^1}\ket{F_-^2}\right)\ket{1}\bigg] \\ &=& \frac{1}{4\sqrt{2}} \bigg[\ket{Y^2}\ket{0} + \ket{Z^2}\ket{1}\bigg].
\end{eqnarray*}
Here \begin{eqnarray*}
	\ket{E_+^n} &=& m_1^n\ket{\psi^+}+m_2^n\ket{\psi^-},\\ 
	\ket{E_-^n} &=& m_1^n\ket{\psi^+}-m_2^n\ket{\psi^-},\\ 
	\ket{F_+^n} &=& m_3^n\ket{\phi^+}+m_4^n\ket{\phi^-},\\
	 \ket{F_-^n} &=& m_3^n\ket{\phi^+}-m_4^n\ket{\phi^-},
\end{eqnarray*}
where the coefficients ${m_k^n}$s depend on which of the four outcomes of the measurement, $\{M_k\}_{k=1}^4$ have clicked in the $n$th round. If $M_k$ is the outcome,  $m_k^n=\sqrt{1+3\lambda}$ and $m_l^n=\sqrt{1- \lambda}$ for $l\neq k$.
Similarly, in the round \(n\), after measuring jointly on the nodes, \(2n\) and \(2n+1\),  the $(2n+2)$-party state produced, denoted by \(2\rightarrow 2 + \widetilde{2n}\), can be represented as
\begin{eqnarray}
\label{eq:EBmaxn}
\ket{\Psi_{k}^{n}}&=&\frac{1}{2^{n}\sqrt{2}}\bigg[\ket{Y^n}\ket{0} + \ket{Z^n}\ket{1}\bigg],
\end{eqnarray}
where
\begin{eqnarray}
\label{eq:EBmaxncoeff}
\ket{Y^n} &=& \bigg[ \ket{Y^{n-1}}\ket{F_+^n}+\ket{Z^{n-1}}\ket{E_-^n}\bigg],\nonumber\\
\ket{Z^n} &=& \bigg[ \ket{Y^{n-1}}\ket{E_+^n}+\ket{Z^{n-1}}\ket{F_-^n}\bigg],\nonumber\\
\text{and }
\ket{Y^1} &=& \bigg[ \ket{0}\ket{F_+^1}+\ket{1}\ket{E_-^1}\bigg],\nonumber\\
\ket{Z^1} &=& \bigg[ \ket{0}\ket{E_+^1}+\ket{1}\ket{F_-^1}\bigg].
\end{eqnarray}
Notice that these relations are quite similar to the one in Eqs. (\ref{recursion}) and (\ref{eq:abcd}) although  \(\ket{\chi}\)s and \(\ket{\xi}\)s are replaced by entangling operators. 
In this situation, some interesting features emerge due to the symmetry of the problem. 
\begin{itemize}
\item After the first round,  GGM, $\mathcal{G}$ depends on $\lambda$, as  shown in Fig. \ref{NMS}.
The maximal eigenvalues required to compute GGM comes from the reduced state,  $\rho_{A_1 A_2}$ with  $\lambda <\lambda_c$ and  from $\rho_{A_2 A_3}$  for $\lambda> \lambda_c$. The maximum eigenvalues for \(\lambda <\lambda_c\) and \(\lambda> \lambda_c\) are respectively 
\begin{eqnarray*}
e_{A_1 A_2}&=&\frac{1}{8}\Bigg[3-\lambda+\sqrt{(1-\lambda)(1+3\lambda)}\\ &+& 2\sqrt{2}\sqrt{(1-\lambda)+(1+\lambda+\sqrt{(1-\lambda)(1+3\lambda)})}\Bigg],\\
e_{A_2 A_3} &=& \frac{1}{4}(1+3\lambda),
\end{eqnarray*}
leading to the critical value of GGM, i.e., $\mathcal{G}_c= 0.25$ at $\lambda=\lambda_c$. It is clearly higher than that obtained from the previous  PB protocol. 

\item Interestingly, in the second and third rounds, the GGM remains same at all values of $\lambda$ (see Fig. \ref{NMS}). 

\item In each round, the probabilities of obtaining any of the outcomes are always equal. 
\end{itemize}


In case of non-maximally entangled state, \(\ket{NME}\), after the round, \(n\), the state looks  similar to the one given in Eq. (\ref{eq:EBmaxn}) with the updated normalization, \(\frac{1}{(2 \sqrt{2})^n} \frac{1}{\sqrt{p_k^n}}\), coefficients, \(\ket{Y^n} \rightarrow \cos z \ket{Y^n}\),    \(\ket{Z^n} \rightarrow \sin z \ket{Z^n}\),   and 
%
\begin{eqnarray}
\ket{Y^1} &=&\cos{z} \bigg[  \sin{z} \ket{0}\ket{F_+^1} + \cos{z} \ket{1}\ket{E_-^1}\bigg], \nonumber\\
\ket{Z^1} &=&  \sin{z} \bigg[ \cos{z} \ket{0}\ket{E_+^1} + \sin{z} \ket{1}\ket{F_-^1}\bigg].
\end{eqnarray}
The observations made for maximally entangled state do not remain valid for \(\ket{NME}\) as well as Haar uniformly generated state, \(\ket{\phi^r}\). Specifically, GGM changes in each round although for moderate values of entanglement content of the state, \(\mathcal{G}\) remains almost same in rounds as shown in Fig. \ref{NMS}. This is due to the fact that the entanglement available to expand multipartite entangled state also gradually increases with rounds which is in sharp contrast with the previous product-based inflation process. 

\begin{figure}[ht]
	\includegraphics[width=0.9\linewidth]{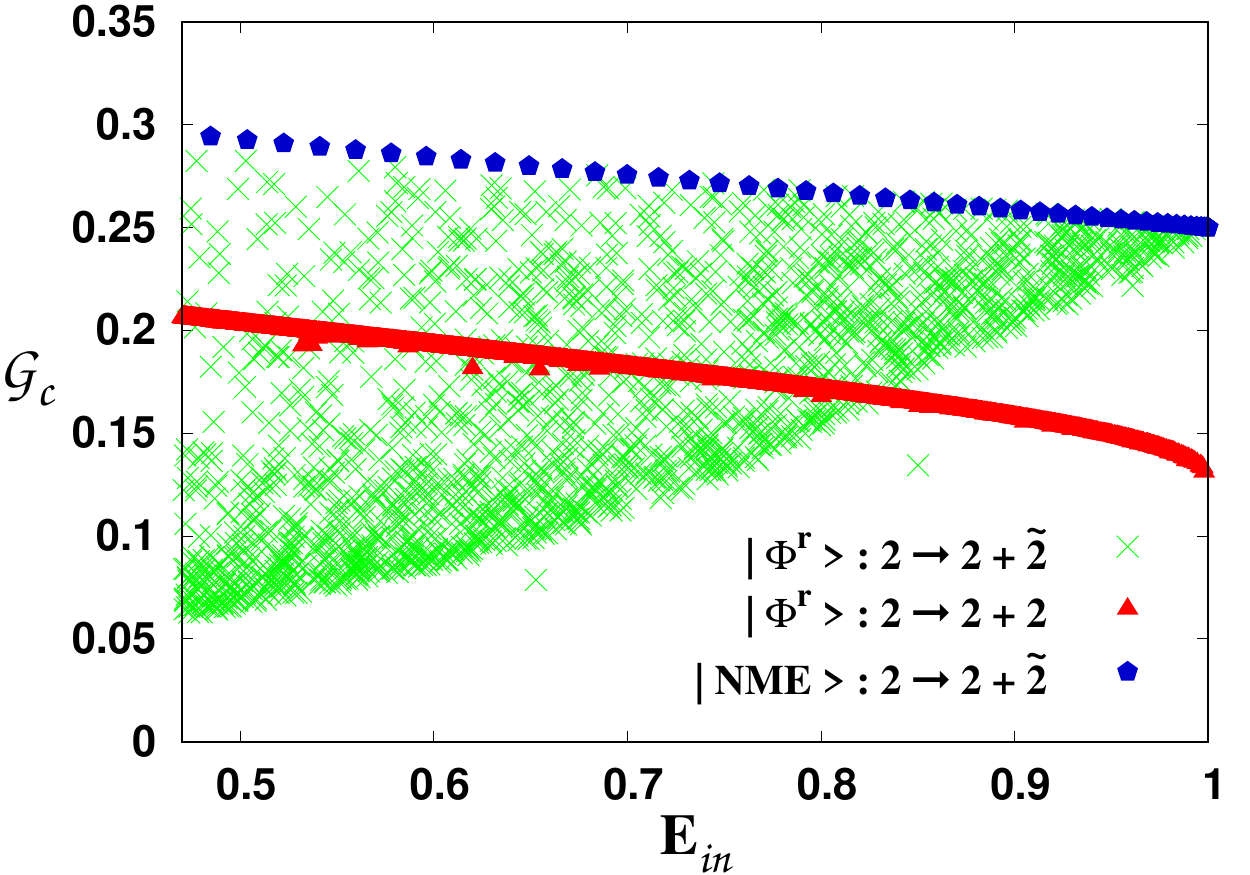}
	\caption{\emph{Entanglement- vs. product-based multipartite inflation procedure. } Critical GGM, $\mathcal{G}_c$ is plotted against initial entanglement, \(E_{in}\) of different resource states. After the first round, the critical GGM is computed for generating states Haar uniformly as well as for NME states (pentagons). The symbols representing PB and EB schemes are respectively triangles and crosses. The number of states simulated in \(5\times 10^3\). Although the vertical axis is   dimensionless, the horizontal axis is in ebits. 
	 }
	\label{com_rand}
\end{figure}

To make the comparison between PB and EB inflation processes more concrete, let us examine  \(\mathcal{G}_c\) for a given initial bipartite entanglement of the resource state, \(E_{in}\),  after the first round of measurement on Haar uniformly generated pure states. Interestingly, it turns out that  \(54.25\%\) of states  creates  less genuine multipartite entanglement  via  the EB protocol than the one that can be achieved by PB scheme  and for the rest, i.e., for \(45.75\%\) of states, EB procedure wins (see Fig. \ref{com_rand}).

%

%
Considering Werner state as initial as well auxiliary states, let us investigate the EB protocol and compare the results with the PB ones. 
For $p\leq\frac{1}{3}$,  negativity monogamy score vanishes for all values of \(\lambda\) as expected.  For $p > \frac{1}{3}$, the variation of negativity monogamy scores reveal the following trends. 
\begin{figure}[ht]
	\includegraphics[width=0.9\linewidth]{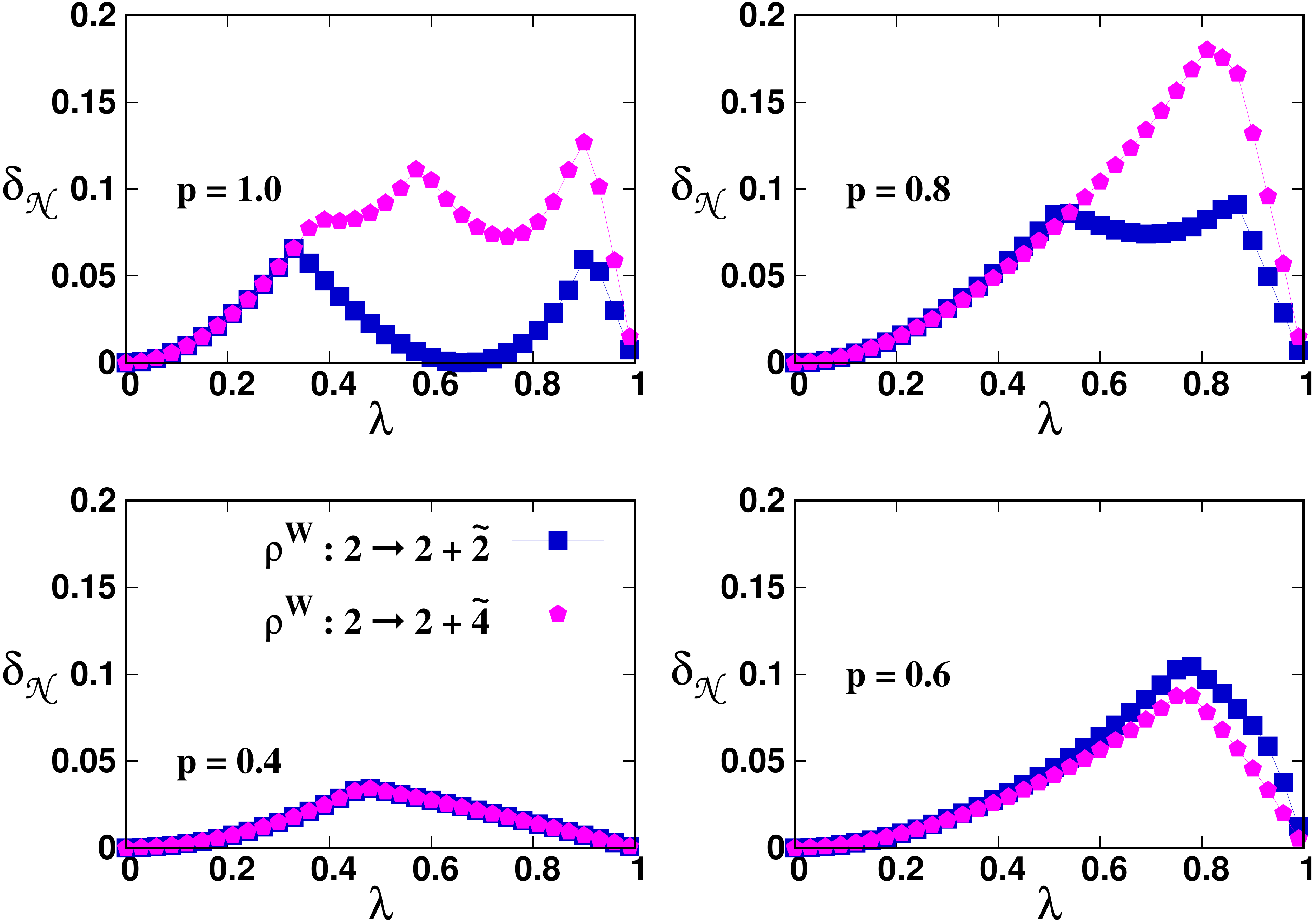}
	\caption{Negativity monogamy scores, $\delta_{\mathcal{N}}$ (vertical axis) vs. $\lambda$ (horizontal axis) with Werner state being the input  as well as auxiliary states. Different \(p\) values are chosen as mentioned in plots. Squares and pentagons represent the first and the second rounds respectively. 
	Both the axes are dimensionless. 
	}
	\label{wernereb}
\end{figure}

\begin{figure}[ht]
	\includegraphics[width=0.9\linewidth]{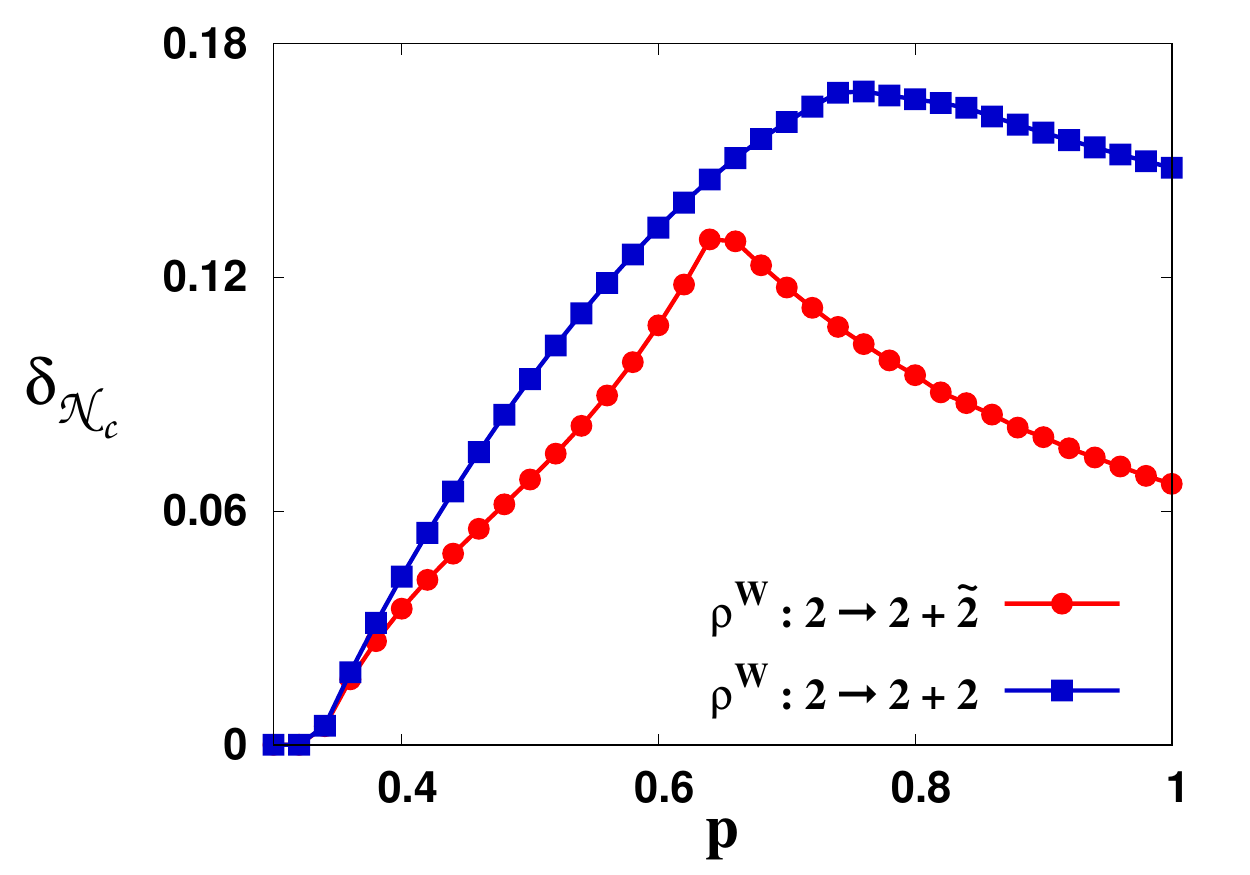}
	\caption{$\delta_{\mathcal{N}_c}$ (ordinate) with respect to noise parameter $p$ (abscissa)  of $\rho^W$. Comparison of  $\delta_{\mathcal{N}_c}$ of the four-party mixed multipartite states obtained in both EB (circles) and PB (squares) protocols from initial resource $\rho^W$ are made. Both the axes are dimensionless.  }
	\label{entvspro}
\end{figure}

\begin{itemize}

\item \emph{Role of noise in EB process.} For a fixed \(\lambda\), the behavior of negativity monogamy score, \(\delta_{\mathcal{N}}\) can be divided into two regions -- (1) when the initial states has less amount of entanglement content, i.e., it is more noisy states with low \(p\), after the first and the second rounds, \(\delta_{\mathcal{N}}\) for the resulting states  almost coincide; (2) when the resource  and the auxiliary states contain substantial amount of entanglement, \(\delta_{\mathcal{N}}\) can be made higher with the increase of rounds as depicted in Fig. \ref{wernereb}, especially near critical value of \(\lambda\). 

\item \emph{PB vs. EB inflation.} Interestingly, \(\delta_{\mathcal{N}_c}\) with PB protocol is always higher than that of the EB protocol for any strength of noise, \(p\) as shown in Fig. \ref{entvspro}.  Notice, however, that for a fixed amount of noise in the channel and for a fixed sharpness parameter, \(\lambda\), EB method can also give advantage than that of the product ones. 
\end{itemize}






\section{Conclusion}
\label{sec:conclu}

Sharing genuine multipartite entangled (GME) states is undoubtedly advantageous for designing several quantum protocols ranging from measurement-based quantum computation to secret key distribution. Over the years, several processes have been developed to create and detect entanglement in shared multipartite systems. Such production schemes include series of single-qubit as well as two-qubit quantum gate operations, projective measurements in a star network, quantum state transfer via  teleportation to name a few. 

In this article, we developed a  mechanism to inflate a genuinely multipartite entangled state with the aid of a single bipartite or multipartite entangled state, several auxiliary systems and controlled unsharp measurement, which we referred to as a product-based  (PB) inflation process.  Notice that instead of unsharp measurement, if we use projective measurement, such multipartite entanglement production is not possible. The successful generation of a multipartite entangled state is guaranteed by measuring multipartite entanglement geometrically, as well as by using monogamy-based measures. Specifically, starting from bipartite pure  states, we determined a recursion relation for obtaining a multipartite state after arbitrary rounds. In the case of  initial bipartite states, we illustrated that for a given sharpness parameter, there is a unique non-maximally entangled state that can create a maximal GME state, thereby showing the importance of sharing non-maximally entangled states over maximally entangled ones. The results were supported also by simulating Haar uniformly generated pure states and for  noisy entangled states, namely the Werner state as inputs. In particular, in the latter situation,  we found that in each round, there is a threshold noise at which multipartite entanglement produced is maximum. It clearly indicates that there is a trade-off relation between the entanglement content of the initial state and the entangling positive operator-valued measurement acted on two parties. We then extend the idea for the situation when the initial shared state is the tripartite state instead of two-party states. Here we showed that after the first round, the states belonging to the W-  and the GHZ-class states  perform equally good  to create  multipartite entangled states. It seems plausible that different initial states along with suitable weak measurements can lead to different classes of multipartite entangled states, which are potential candidates for quantum information processing tasks.

Instead of single-qubit auxiliary states, if we use several copies of the initial entangled states as the starting point, the unsharp measurements can again create a highly multipartite entangled state, which we call  entanglement-based (EB)  inflation scheme. Surprisingly, we manifested that although the resource in this process is monotonically increasing with rounds,  there exist certain percentages of pure as well as mixed states, for which the PB method can produce higher GME states compared to that of EB ones.  The entire analysis reveals that there is a competition between  entanglement in inputs and the entangling measurement operators which requires much more careful analysis in different information processing tasks.

\section*{Appendix A: Multipartite Entanglement Quantifier }

A multipartite pure state is said to be genuinely multiparty entangled if it is not product in any bipartition. Genuine multipartite entanglement can be quantified by using the geometry of quantum states. In particular, generalized geometric measure (GGM), \(\mathcal{G}\),  of a given pure state, \(|\psi_N\rangle\) is the minimum distance between the given state with the set of non-genuinely multipartite entangled states, denoted by \(nG\). By using Schmidt decomposition in each bipartition \cite{sen'10, biswas'14}, it reduces to 
\(\mathcal{G} (|\psi_N\rangle) = 1 - \max_{\{|\phi \rangle \in nG\}} |\langle \phi | \psi_N \rangle |^2 = 1 - \max (\{\lambda^m_{A:B})\}\), where \(\lambda^m_{A:B}\) is the maximum Schmidt coefficients in all the non-trivial bipartitions of \(|\psi_N\rangle\). Although the measure can be computed easily for  pure states with reasonable number of parties,  it is hard to get a closed form for mixed states. 

We use negativity monogamy score to measure entanglement distribution in multipartite mixed state \cite{coffman'00, dhar'17}. It is defined for arbitrary density matrix, \(\rho_{12\ldots N}\), as \(\delta_{\mathcal{N}} = \mathcal{N} (\rho_{1:2\ldots N}) - \sum_{i=2}^N \mathcal{N} (\rho_{1i})\), where \(\mathcal{N} (\rho_{1:2\ldots N})\) is the   negativity in the bipartition of \(1:2 \ldots N\), and  \(\mathcal{N} (\rho_{1i})\)s are the reduced density matrices of  \(\rho_{12\ldots N}\). Here, for two-qubit states, negativity, \(\mathcal{N} (\rho_{12} )\) \cite{zyckohoro'98, zyczkowski'99, vidal'02, plenio'05} reduces to the absolute value of a negative eigenvalue of the partial transposed state with respect to one of parties  \cite{peres'96, horodecki'96} while logarithmic negativity is defined as \(E_{\mathcal{N}}(\rho_{12}) = \log_2 (2 \mathcal{N}(\rho_{12}) +1) \). 
\vspace{0.2 cm}

\section*{Appendix B: Probabilities of the outcome in inflation process}

When the shared state is the maximally entangled state and the qubits auxiliary systems are used for the expansion of multipartite entanglement, 
we now discuss the structure of the  probabilities to obtain the specific outcome  at each round.  In the first round, each of the four measurements has the same probability, $p_k^1=\frac{1}{4}$, independent of  $\theta_1,$  $\phi_1$. 

 When the outcome is \(\sqrt{M_1}\) both in the second and the third rounds,  the corresponding probabilities which depend on the auxiliary state parameters  read as
\begin{eqnarray}
\label{p2}
\nonumber p_1^2 &=& \frac{1}{p_1^1}\times \frac{1}{64}\bigg[4-2\lambda\left(1-\lambda+\sqrt{(1-\lambda)(1+3\lambda)}\right)\\ 
&\times&\left( \cos\theta_1 \cos\theta_2 - \cos(\phi_1-\phi_2)\sin\theta_1 \sin\theta_2 \right) \bigg],
\end{eqnarray}
and 
\begin{eqnarray*}
p_1^3 &=& \frac{1}{p_1^2}\times \frac{1}{64}\bigg[4-2\lambda\left(1-\lambda+\sqrt{(1-\lambda)(1+3\lambda)}\right)\\ 
&\times & ( \cos\theta_1 \cos\theta_2 + \cos\theta_2 \cos\theta_3  - \cos(\phi_1-\phi_2) \sin\theta_1 \sin\theta_2 )\\ 
&+& 2\lambda\left(\lambda(1-\lambda)+\sqrt{(1-\lambda)(1+3\lambda)} \right)\\
&\times& (\cos(\phi_2-\phi_3)\sin\theta_2 sin\theta_3) \\ 
&+& 4\lambda^2 (1-\lambda)(\cos\theta_1 \cos\theta_3 + \lambda \cos(\phi_1-\phi_3)\sin\theta_1 sin\theta_3) \bigg].
\end{eqnarray*}
Similarly, if  $\sqrt{M_1}$ clicks successively in all the \(n\) rounds, the probability of clicking it reduces to
\begin{eqnarray}
p_1^n=\frac{\bra{Z_1^n}\ket{Z_1^n}}{p_1^{n-1}},
\end{eqnarray}
where $\ket{Z_k^n}$ is given in Eq. (\ref{nth_state}).
Therefore, one can compute the probabilities of obtaining any  outcome in a specific round based on the outcomes of the previous rounds via the recursion relation.


%

\section*{Acknowledgement}
We acknowledge the support from Interdisciplinary Cyber Physical Systems (ICPS) program of the Department of Science and Technology (DST), India, Grant No.: DST/ICPS/QuST/Theme- 1/2019/23 and SM acknowledges Ministry of Science and Technology in Taiwan (Grant no. 110-2811-M-006 -501).  We  acknowledge the use of \href{https://github.com/titaschanda/QIClib}{QIClib} -- a modern C++ library for general purpose quantum information processing and quantum computing Ref. \cite{titas} and cluster computing facility at Harish-Chandra Research Institute.


\begin{thebibliography}{99}

\bibitem{hans'01} R. Raussendorf and H.-J. Briegel, Phys. Rev. Lett. {\bf 86}, 5188 (2001).



\bibitem{qcomp1} M. Hein, J. Eisert, and H. J. Briegel, Phys. Rev. A {\bf 69}, 062311 (2004).
\bibitem{qcomp2}  M. Hein, W. D{\"u}r, J. Eisert, R. Raussendorf, M. Van den Nest, and H. J. Briegel, in Proceedings of the International School of Physics “Enrico Fermi” on “Quantum Computers, Algorithms and Chaos” (2006), arXiv:quant-ph/0602096.

\bibitem{qcomp3} H. J. Briegel, D. E. Browne, W.D{\"u}r, R. Raussendorf, and M. Vanden Nest, Nat. Phys. {\bf 5}, 19 (2009).

\bibitem{beals'13} R. Beals, S. Brierley, O. Gray, A.W. Harrow, S. Kutin, N. Linden, D. Shepherd and M. Stather,  Proc. R. Soc. London A {\bf 469}, 20120686 (2013).

\bibitem{tele} C. H. Bennett, G. Brassard, C. Crepeau, R. Jozsa, A. Peres, and W. K. Wootters, Phys. Rev. Lett. {\bf 70}, 1895 (1993).
\bibitem{teleexp} D. Bouwmeester, J.-W. Pan, K. Mattle, M. Eibl, H. Weinfurter, and A. Zeilinger, Nature {\bf 390}, 575 (1997).
\bibitem{tele1}  M. Murao, D. Jonathan, M. B. Plenio, and V. Vedral, Phys. Rev. A {\bf 59}, 156 (1999).
\bibitem{tele2}  A. Grudka, Acta Phys. Slov. {\bf 54}, 291 (2004), arXiv:quant-ph/0303112.
\bibitem{tele3} A. Sen(De) and U. Sen, Phys. Rev. A {\bf 81}, 012308 (2010).

\bibitem{dc}  C. H. Bennett and S. J. Wiesner, Phys. Rev. Lett. {\bf 69}, 2881 (1992).
\bibitem{dcexp}  K. Mattle, H. Weinfurter, P. G. Kwiat, and A. Zeilinger, Phys. Rev. Lett. {\bf 76}, 4656 (1996).
\bibitem{dc1} A. Sen (De) and U.Sen, Phys. News {\bf 40}, 17 (2011) (arXiv:1105.2412).
\bibitem{dc2}  D. Bru{\ss}, G. M. D’Ariano, M. Lewenstein, C. Macchiavello, A. Sen(De), and U. Sen, Phys. Rev. Lett. {\bf 93}, 210501 (2004).

\bibitem{dc3} D. Bru{\ss}, M. Lewenstein, A. Sen(De), U. Sen, G. M. D’Ariano, and C. Macchiavello, Int. J. Quant. Info. {\bf 4}, 415 (2006). 
\bibitem{dc4} M. Horodecki and M. Piani, J. Phys. A: Math. Theor. {\bf 45}, 105306 (2012).
 \bibitem{dc5} Z. Shadman, H. Kampermann, D. Bru{\ss}, and C. Macchiavello, Phys. Rev. A {\bf 85}, 052306 (2012). 
 \bibitem{dc6} T. Das, R. Prabhu, A. Sen(De), and U. Sen, Phys. Rev. A {\bf 90}, 022319 (2014); Phys. Rev. A {\bf 92}, 052330 (2015).
 
\bibitem{ekert'91} A. K. Ekert, Phys. Rev. Lett. {\bf 67}, 661 (1991).
 \bibitem{key1} T. Jennewein, C. Simon, G. Weihs, H. Weinfurter, and A. Zeilinger, Phys. Rev. Lett. {\bf 84}, 4729 (2000).
 \bibitem{key2} N. Gisin, G. Ribordy, W. Tittel, and H. Zbinden, Rev. Mod. Phys. {\bf 74}, 145 (2002).
\bibitem{key3} M. Hillery, V. Buzek, and A. Berthiaume, Phys. Rev. A {\bf 59}, 1829 (1999).

\bibitem{key4}  R. Cleve, D. Gottesman, and H.-K. Lo, Phys. Rev. Lett. {\bf 83}, 648 (1999).
 \bibitem{key5} A. Karlsson, M. Koashi, and N. Imoto, Phys. Rev. A {\bf 59}, 162 (1999). 
 \bibitem{hillery'06}M. Hillery, M. Ziman, V. Buzek, M. Bielikova, Phys. Lett. A {\bf 349}, 75 (2006).

\bibitem{xu'14} G.-B. Xu, Q.-Y. Wen, F. Gao, S.-J. Qin, Quantum Information Processing {\bf 13}, 2587 (2014).

 
 



\bibitem{qmanybody} M. Lewenstein, A. Sanpera, V. Ahufinger, B. Damski, A. Sen(De), and U. Sen, Adv.  Phys. {\bf 56}, 243 (2007).

\bibitem{qmanybody1} L. Amico, R. Fazio, A. Osterloh, and V. Vedral, Rev. Mod. Phys. {\bf 80}, 517 (2008).

\bibitem{qmanybody2} G. D. Chiara and A. Sanpera, Rep. Prog. Phys. {\bf 81}, 074002 (2018).

\bibitem{dqpt}  M. Heyl, Rep. Prog. Phys. {\bf 81}, 054001 (2018).

\bibitem{qmanybody3}  S. Haldar, S. Roy, T. Chanda, A. Sen(De), and U. Sen, Phys. Rev. B {\bf 101}, 224304 (2020).




\bibitem{linden'99} N. Linden, S. Popescu, B. Schumacher, M. Westmoreland, 	arXiv:quant-ph/9912039 (1999).

\bibitem{pirker'18} A. Pirker, J. Wallnofer, and W. Dur, New J. Phys. {\bf 20}, 053054 (2018).

\bibitem{jaksch'99} D. Jaksch, H. J. Briegel, J. I. Cirac, C. W. Gardiner, P.  Zoller, Phys. Rev. Lett. {\bf 82}, 1975 (1999).



\bibitem{hans'98} H.-J. Briegel, W. Dur, J. I. Cirac, and P. Zoller, Phys. Rev. Lett. {\bf 81}, 5932 (1998).

\bibitem{zukowski'93} M. Zukowski, A. Zeilinger, M. A. Horne, and A. K. Ekert, Phys. Rev. Lett. {\bf 71}, 4287 (1993).
\bibitem{swapping2}  M. {\.Z}ukowski, A. Zeilinger, and H. Weinfurter, Annals N.Y. Acad. Sci. {\bf 755}, 91 (1995).
\bibitem{swapping3} S. Bose, V. Vedral, and P.L. Knight, Phys. Rev. A {\bf 57}, 822 (1998).
\bibitem{swapping4}  S. Bose, V. Vedral, and P.L. Knight, Phys. Rev. A {\bf 60}, 194 (1999).

\bibitem{bennett'96} C. H. Bennett, G. Brassard, S. Popescu, B. Schumacher, J. A. Smolin, and W. K. Wootters, Phys. Rev. Lett. {\bf 76}, 722 (1996).



\bibitem{epping'16} M. Epping, H. Kampermann, D. Bruss, New Journal of Physics {\bf 18}, 053036 (2016).


\bibitem{swapping5}   A. Sen (De), U. Sen, C. Brukner, V. Buzek, and M. {\.Z}ukowski, Phys. Rev. A {\bf 72}, 042310 (2005).

\bibitem{network2} D. Cavalcanti, M.L. Almeida, V. Scarani, and A. Acin, Nat. Commun. {\bf 2}, 184 (2011).

\bibitem{swapping6} R. Banerjee, S. Ghosh, S. Mal, and A. Sen(De), Phys. Rev. Research {\bf 2}, 043355 2020). 

\bibitem{fusion} P. Walther, K .J.  Resch, and A. Zeilinger, Phys. Rev. Lett. {\bf 94}, 240501 (2005).


\bibitem{lee'05} J.-S. Lee, A. K. Khitrin, Phys. Rev. Lett. {\bf 94}, 150504 (2005).


\bibitem{fusion1} S. K. {\"O}zdemir,  E Matsunaga, T Tashima, T. Yamamoto, M. Koashi and N. Imoto, New J. Phys. {\bf 13}, 103003 (2011).

\bibitem{fusion2} X.-P. Zang, M.Yang, F. Ozaydin, W. Song and  Z.-L. Cao,  Sci. Reports {\bf 5}, 16245 (2015).

\bibitem{unsharp1} P. Busch, P. J. Lahti, and P. Mittelstaedt, \emph{The Quantum Theory of Measurement}, Springer: Berlin, Germany, (1996).
\bibitem{unsharp1a}   P. Busch, Phys. Rev. D {\bf 33}, 2253 (1986).

\bibitem{adv1}  R. Derka, V. Buzek, and A. K. Ekert, Phys. Rev. Lett. {\bf 80}, 1571 (1998). 
\bibitem{adv2}  J. Shang, A. Asadian, H. Zhu, O. G{\"u}hne,  Phys. Rev. A {\bf 98}, 022309 (2018).
\bibitem{adv3}   D. Dieks, Phys. Lett. A {\bf 126}, 303 (1988).
\bibitem{adv4} A. Peres, Phys. Lett. A {\bf 128}, 19 (1988).
\bibitem{adv5} T. Vertesi and E. Bene,  Phys. Rev. A {\bf 82}, 062115 (2010).
 \bibitem{adv6} S. Gomez, A. Mattar, E. S. Gomez, D. Cavalcanti, O. Jimenez Farias, A. Acin, and G. Lima,  Phys. Rev. A {\bf 97}, 040102(R) (2018).





\bibitem{unsharp4} S. Roy, A. Bera, S. Mal, A. Sen(De), and U. Sen, Phys. Letts. A {\bf 392},
127143 (2021).

\bibitem{unsharp5} C. Srivastava,  S. Mal, A. Sen(De), and U. Sen, Phys. Rev. A {\bf 103},  032408 (2021).


\bibitem{biswas'14} A. Biswas, R. Prabhu, A. Sen(De), and U. Sen, Phys. Rev.
A {\bf 90}, 032301 (2014).

\bibitem{dhar'17} H. Dhar, A. Pal, D. Rakshit, A. Sen(De), and U. Sen,
Monogamy of Quantum Correlations - A Review. In: Lectures on General Quantum Correlations and their Applications. Quantum Science and Technology. (Springer, Cham,
2017).

\bibitem{bengtsson'06} I. Bengtsson and K. {\.Z}yczkowski, Geometry of Quantum
States (Cambridge University Press, 2006).

\bibitem{werner'89} R.F. Werner, Phys. Rev. A {\bf 40}, 4277 (1989).

\bibitem{GHZ} D.M.  Greenberger,  M.A.  Horne,  and  A.  Zeilinger, \emph{ in
Bell’s Theorem and the Conceptions of the Universe},  edited  by  M.
Kafatos (Kluwer Academic, Dordrecht, 1989).

\bibitem{dur'00} W. Dür, G. Vidal, and J. I. Cirac, Phys. Rev. A {\bf 62}, 062314
(2000).

\bibitem{coffman'00} V. Coffman, J. Kundu, and W. K. Wootters, Phys. Rev. A
{\bf 61}, 052306 (2000).

\bibitem{hill'97} S. Hill and W. K. Wootters, Phys. Rev. Lett. {\bf 78}, 5022
(1997).
\bibitem{wootters'98} W. K. Wootters, Phys. Rev. Lett. {\bf 80}, 2245 (1998).


\bibitem{Pritamnew} P. Halder, S. Mal, and A. Sen (De),(unpublished). 




\bibitem{zyckohoro'98} K. {\.Z}yczkowski, P. Horodecki, A. Sanpera, and M. Lewenstein, Phys. Rev. A {\bf 58}, 883 (1998).

\bibitem{zyczkowski'99} K. {\.Z}yczkowski, Phys. Rev. A {\bf 60}, 3496 (1999).

\bibitem{vidal'02} G. Vidal and R. F. Werner, Phys. Rev. A {\bf 65}, 032314 (2002).

\bibitem{plenio'05} M. B. Plenio, Phys. Rev. Lett. {\bf 95}, 090503 (2005).

\bibitem{vn} The von-Neumann entropy of a quantum state, \(\sigma\) is defined as \(S(\sigma) = - \mbox{Tr} \sigma \log_2 \sigma\). 

\bibitem{rivu'20}  R. Gupta, S. Gupta, S. Mal, and A. Sen (De), arXiv:2005.04009.














\bibitem{soorya'19} S. Rethinasamy, S. Roy, T. Chanda, A. Sen(De), and U. Sen, Phys. Rev. A {\bf 99}, 042302 (2019).

\bibitem{Wstate1} A. Sen(De), U. Sen, M. Wiesniak, D. Kaszlikowski, and M. {\.Z}ukowski, Phys. Rev. A {\bf 68}, 062306 (2003). 
\bibitem{Wstate2} D. Kaszlikowski, A. Sen(De), U. Sen, V. Vedral, and A. Winter, Phys. Rev. Lett. {\bf 101}, 070502 (2008).
\bibitem{Wstate3} T. J. Barnea, G.P{\"u}tz, J. B. Brask, N. Brunner, N. Gisin, and Y.-C. Liang, Phys. Rev. A {\bf 91}, 032108(2015).
\bibitem{Wstate4} W. Laskowski, T. Vertesi, and M. Wiesniak,  J. Phys. A: Math. Theor. {\bf 48}, 465301 (2015).
\bibitem{Wstate5} S. Roy, T. Chanda, T. Das, A. Sen(De), and U. Sen, Phys. Lett. A {\bf   382}, 1709 (2018).

\bibitem{titas} \url{http://titaschanda.github.io/QIClib}

\bibitem{peres'96} A. Peres, Phys. Rev. Lett. {\bf 77}, 1413 (1996).
\bibitem{horodecki'96} M. Horodecki, P.Horodecki, and R. Horodecki, Phys. Lett. A {\bf 223}, 1 (1996).



\end{thebibliography}
\end{document}